\documentclass[aps,13pt,eqsecnum, preprint,nofootinbib,superscriptaddress]{revtex4}
\usepackage{amssymb,amsmath,amsthm,graphicx,amscd}
\usepackage{enumerate,color,changepage,ulem,bm,subcaption}
\usepackage[hidelinks]{hyperref}

\begin{document}

\title{Quantum Parametric Oscillator Heat Engines in Squeezed Thermal Baths: Foundational Theoretical Issues}

\author{Onat Ar{\i}soy}
\email{oarisoy@umd.edu}
\affiliation{Chemical Physics Program and Institute for Physical Science and Technology,  University of Maryland, College Park, Maryland 20742, USA}
\author{Jen-Tsung Hsiang}
\email{cosmology@gmail.com}
\affiliation{Center for High Energy and High Field Physics, National Central University, Chungli 32001, Taiwan, ROC}
\author{Bei-Lok Hu}
\email{blhu@umd.edu}
\affiliation{Joint Quantum Institute and Maryland Center for Fundamental Physics,  University of Maryland, College Park, Maryland 20742, USA}

%
%
%

\begin{abstract}
In this paper we examine some  foundational issues of a class of quantum engines where the system consists of  a single quantum parametric  oscillator,  operating in an Otto cycle  consisting  of 4 stages of two alternating phases:  the isentropic phase is detached from any bath (thus a closed system) where the natural frequency of the oscillator is changed from one value to another, and the isothermal phase where the system (now rendered open) is put in contact with one or two squeezed  baths of different temperatures, whose nonequilibrium dynamics follows the Hu-Paz-Zhang (HPZ) master equation for quantum Brownian motion. The HPZ equation is an exact  nonMarkovian equation which preserves the positivity of the density operator and is valid for a) all temperatures, b) arbitrary spectral density of the bath, and c) arbitrary coupling strength between the system and the bath. 
Taking advantage of these properties we examine some key foundational issues of theories of quantum open and squeezed systems for these two phases of the quantum Otto engines.  This include, i) the nonMarkovian regimes for non-Ohmic, low temperature baths, ii) what to expect in nonadiabatic frequency modulations,  iii) strong system-bath coupling, as well as iv) the proper junction conditions between these two phases.  Our aim here is not to present ways for attaining higher efficiency but to build a more solid theoretical foundation for quantum engines of continuous variables covering a broader range of parameter spaces hopefully of use for exploring such possibilities.   
\end{abstract}

\maketitle

\hypersetup{linktoc=all}
\setcounter{tocdepth}{2}

\clearpage
\baselineskip=18pt
\newtheorem{theorem}{Theorem}[section]
\newtheorem{corollary}[theorem]{Corollary}
\newtheorem{lemma}[theorem]{Lemma}
\newtheorem{example}[theorem]{Example}
\newtheorem{proposition}[theorem]{Proposition}
\numberwithin{equation}{section}
\allowdisplaybreaks

\section{Introduction}

In the broad range of applications of classical thermodynamics the subject of thermal engines and refrigerators certainly stands out,  not only in terms of its practical application utility, but also in view of their foundational theoretical values. Thermodynamic laws are often phrased in terms of engine or refrigerator efficiency: ``No engine can have a higher efficiency than that of an ideal Carnot engine" (Second Law); 
``No refrigerator can cool a physical object to the absolute zero temperature" (Third Law).  In the new field of quantum thermodynamics \cite{QTDbooks} the subject of \textit{quantum} engines and refrigerators \cite{KosLev} occupy an equally important place. One can ask, e.g., how would the Third Law be phrased  in terms of a quantum refrigerator \cite{Alicki}? Quantum nature takes prominence beyond  the validity ranges of classical thermodynamics in \textit{small systems} \cite{small}, at \textit{low temperatures}, with \textit{strong coupling} between the system and its environment \cite{MarPaz,Uzdin} and in regimes where \textit{non-Markovian} behaviors associated with quantum memory and correlations (see, e.g., \cite{Bre16,DeV17,HuePle}) can no longer be ignored.  Add to this list the increased weight of \textit{quantum fluctuations} and under \textit{out-of-equilibrium conditions} (see, e.g., \cite{QTD1} and references therein) -- for instance, there are claims that heat engines may be operated beyond the Carnot limit by exploiting stationary, nonequilibrium reservoirs \cite{Klaers}. 

The exploration and exploitation of these new elements, new regimes and new processes  hold great promises for advancing quantum sciences and engineering involving the thermodynamics of such systems.  However, the familiar paradigms, tools and concepts in traditional equilibrium thermodynamics, mean field dynamics and linear response theories cannot serve these purposes well. Fortunately, two related families of powerful theories, one for open and dissipative quantum systems \cite{UW12,BrePet,RivHue} and the other for nonequilibrium quantum fields \cite{CalHu88,CalHu08,JR09,AK11,Berges}, which have been in development over the last three decades, are pretty well established now to  meet these new challenges. 

\paragraph{Otto Engines}

Quantum engine is a topic of active current interest, e.g., a quantum engine made up of a single calcium ion in a Paul trap operating between a cold reservoir of laser-cooling beam and a hot reservoir of electric field noise \cite{Ross,Abah} has received much recent attention.  We refer to the lucid  reviews by Kosloff \& co-workers \cite{KosLev,KosRez} and the references cited therein  for a general background to  quantum heat engines and refrigerators (see also  \cite{DeffLutz,AbahLutz,Manzano,Klaers,Reid} before the 2017 review and \cite{Niedensu,Pezzutto} with those in \cite{ColMod} thereafter). In this paper we examine a class of quantum Otto engines with  continuous system  variables (in contrast to discrete systems such as spins or qubits) interacting with a bath also of continuous variables (in contrast to stochastic actions such as in the collision models). Our system is a  quantum parametric oscillator (with time-dependent natural frequency)  which in principle can interact with two   squeezed baths at different temperatures,  operating under fully nonequilibrium conditions.  In the simpler setup we shall analyze here,   the parametric oscillator enter in one (isentropic) phase without any bath,  and in another (isothermal) phase the oscillator's frequency does not change, but it interacts with a (fixed valued) squeezed thermal bath. 

The classical Otto cycle consists of four stages made up of two alternating phases: the adiabatic (no heat exchange) phase and the isochoric  (equal volume) phase.
(We use the same labeling and ordering as Fig 1b of \cite{Manzano})  We shall use the terminology `isentropic'  (no heat exchange) for the former,  reserving the term adiabatic and nonadiabatic to describe how fast the natural frequency of the system oscillator is modulated,  and `isothermal' for the latter, as  the system will be brought in contact with a constant temperature heat bath. In the isentropic phase  a modulation in the frequency of the system oscillator is implemented, but  since the system is thermally insulated from its environment and its evolution is unitary, we shall view it as a closed system (i.e., not accounting for an external drive which alters the oscillator frequency).    This phase is sometimes called the `power phase'. In the isothermal phases the system can be placed in contact with  two baths of different squeezing parameters.  (In the simpler setup we'll analyze only one bath is squeezed.)  As such the standard processes in open systems occur, like relaxation, dissipation, decoherence etc. These phases are sometimes called the `relaxation phase' of the cycle, but of course there are more activities than just relaxation.  The gain in efficiency is explained in the review papers referred to above.

\paragraph{Theoretical Background} 

Many factors are involved in these two phases but for reasons  elaborated below we shall focus on a selected few of foundational theoretical interest. By `foundational' we are referring to the theories of open quantum systems  and of squeezing  quantum systems in the two different phases of the Otto cycle.  We first explain the reasons why  for the description of the open system phases  the commonly invoked Lindblad  or the Caldeira-Leggett master equations are inadequate for treating the very low temperature regimes. We then introduce the route we take here based on the Hu-Paz-Zhang (HPZ) master equation for quantum Brownian motion (QBM). The HPZ equation is an exact  nonMarkovian equation which preserves the positivity of the density operator and is valid for a) all temperatures, b) arbitrary spectral density of the bath, and c) arbitrary coupling strength between the system and the bath. The non-Ohmic, low temperature regime is where nonMarkovian behavior gets pronounced.   

Using this genric model and taking advantage of these attractive properties we shall examine the theoretical foundation of  quantum engines based on the quantum Brownian motion in aspects pertaining to  1) the {\it very low temperature} regime,  2) the {\it late time squeezed thermal state}, correcting some flaws  in the literature,  3) the {\it strong system-bath coupling}  in the isothermal phases, as well as  4) the {\it nonadiabatic frequency variation} in the isentropic phases, and finally 5) the  {\it junction conditions} between the isentropic and the isothermal phases.  We emphasize that the aim of this work is not to present ways for attaining higher efficiency in the sense of `you can do this for better results',  but to provide a more solid and broader foundation for different aspects of the theories involved.  For example, we shall include nonMarkovian behavior, strong coupling  and nonadiabatic drives in our discussions but make clear that at the present level of theoretical understanding, we can only work with weak coupling in the engine design, likewise for adiabatic drives,  since they yield higher efficiency.  The purposes for our foundational theoretical studies are twofold: to scrutinize the validity of assumptions and approximations in the existing theories so mistakes can be avoided, and to prepare a broader base for future searches into wider parameter spaces for more efficient engines or other quantum thermodynamic  applications.        


\subsection{Theories based on Markovian Master Equations}

Quantum heat engines operating at sufficiently high temperatures are relatively easy to construct, and the theories supporting them are well understood. See,  e.g., \cite{Abah,AbahLutz}.  The challenge is  for low temperatures.  Many theoretical work in quantum heat engines invoke the `collision model' for the baths \cite{ColMod}. For instance, the claimed pride of the nice and thorough work of Ref \cite{Manzano} is the  derivation of the efficiency of a quantum  Otto engine at \textit{low temperatures}. The authors justified the use of the Lindblad equation with a collision model derivation. At sufficiently high temperatures this is justified, but we have serious reservations in the adequacy of Markovian collision models to describe very low temperature phenomena. In its original context, for the description of the dynamics of gas molecules, single-hit collision assumption yields good results primarily for dilute gases at high temperatures. 

Markovian processes are aptly described by Markovian master equations, such as the very popular Lindblad equation. The question is, can one still use this model for the opposite regime of very low temperatures or very high  density -- when correlations build up in collisions and ignoring the memory becomes untenable? We only raise the issue here but save the discussion to a sequel paper \cite{AHH2} where we shall examine the collision models, both Markovian and nonMarkovian, and analyze them in relation to the corresponding quantum Brownian models where a reliable master equation is indeed available for the study of nonMarkovian effects.   

Another important Markovian master equation is the Caldeira-Leggett  (CL) master equation for quantum Brownian motion. Unfortunately both the Lindblad  and the CL  master equations have problems at low temperatures: Lindblad violates the uncertainty principle while CL  loses the positive definiteness of the density matrix.
This is an important point which needs some elaboration.

\paragraph{Lindblad-GKS and Caldeira-Leggett master equations are problematic at very low temperatures}

 Many theoretical models of quantum engines  use the Lindblad \cite{Lindblad}-Gorini-Kossakowski-Sudarshan (GKS) master equation \cite{GKS} because  1) it is mathematically sound, having a completely positive (CP) dynamical semigroup property 2) it is commensurate with the Born-Markov master equation \cite{BornMark} often used in quantum optics and 3) it is consistent with the Caldeira-Leggett (CL) master equation \cite{CalLeg} for a Brownian motion model valid with Ohmic bath at high temperatures -- note the CL master equation for the density matrix is not positive-definite \cite{AmbBun}. All is well until one explores its low temperature behavior. 

As pointed out by Haake \cite{Haake}, Garraway \cite{Garraway} and many others, \textit{there are serious problems of the Lindblad master equation at very low temperatures}.  For example, 
it leads to  a violation of the Heisenberg uncertainty principle. There are proposals to ameliorate these maladies, e.g.,  in Ref \cite{Maciej} a term is added to the Born-Markov master equation, which vanishes in the classical limit, bringing the equation to the Lindblad form and satisfying the uncertainty principle. In  Ref \cite{Jacobs} a Lindblad-form master equation for \textit{weakly-damped} systems  accurate for all regimes is derived, which can serve as a replacement for the Bloch-Redfield equation for thermal damping that is completely positive. 
These  amendments target  one kind of deficiency but remain restricted by other limitations or approximations. 

Thus the focus of attention would be to find a master equation,  for the particular class of problems one wants to solve, which is both mathematically correct and physically sound at  low temperatures, the regime where nonMarkovian behaviors become important.


\subsection{QBM nonMarkovian master equation valid for non-Ohmic baths at low temperatures}

For systems and baths which can suitably be modeled by quantum harmonic oscillators  there is such an equation, the Hu-Paz-Zhang (HPZ) master equation for quantum Brownian motion \cite{HPZ92,HPZ93}.  (An equivalent representation is via the Fokker-Planck equation governing the reduced Wigner function \cite{HalYu96}.)  It is an exact  nonMarkovian master equation which preserves the positivity of the density operator and is valid for   all temperatures,  arbitrary spectral density of the bath and  arbitrary coupling strength between the system and the bath. This offers a much broader range of utilities such as for strong couplings and  for  non-Ohmic baths at ultra-low temperatures, where nonMarkovian behavior are pronounced.   Since it has both the  sound mathematical (positive definite, unlike the CL eqn) and  physical properties (respecting the uncertainty principles, unlike the Lindblad) at low temperatures,  the  HPZ equation can also serve as a good benchmark  where results obtained from other trial models can be compared.  E.g., we will explore the viability of the popular nonMarkovian collision models at low temperatures in our sequel paper \cite{AHH2}.

\subsection{Master equation for QBM of parametric oscillators} 

The Hu-Paz-Zhang master equation \cite{HPZ92,HPZ93} is for a system of one quantum harmonic oscillator of fixed (time-independent) frequency interacting with a bath of harmonic oscillators also of fixed frequencies.  
For the quantum engine we are interested,  in the isentropic phases, the system oscillator frequency  varies with time (parametric oscillators), but without contact with any bath.  In the isothermal phases the system oscillator has a fixed frequency but it interacts with a squeezed thermal bath. A more general theory which encompasses both aspects  exists. 
The HPZ master equation for a parametric quantum oscillator  in a general squeezed thermal bath valid for all spectral densities,  at all temperatures,  and  for arbitrary coupling strength  between the system and its environment was derived by Hu and Matacz  \cite{HM94} using the same method as in the derivation of the original HPZ equation for the Brownian  motion of quantum oscillators.  It was used to demonstrate  the more versatile kinematic approach in the derivation of the Unruh and Hawking effects and going beyond for situations in  more general nonequilibrium conditions \cite{RHK97}.   Cosmological particle creation as squeezing was discussed in \cite{HKM94} and entropy generation in squeezed quantum open systems in \cite{KMH97}.  (For representative work along similar lines of reasoning, see references in e.g., \cite{HHCosSq}).

The nonMarkovian HPZ master equation for parametric oscillators in a squeezed thermal bath \cite{HM94} is an example of a theory with a broader base and more solid foundation we referred to earlier.  Besides nonOhmic baths at low temperatures,  its validity for arbitrary couplings between the system and the bath enables one to examine the coupling strength dependence. We shall show the marked differences between strong and weak coupling.  However, for strong coupling there is some arbitrariness in what portion of the interaction energy should be considered as internal energy there may be some intrinsic ambiguity in how heat is defined.  While this issue is under investigation \cite{HHHMF}, we will return to the practical level when evaluting the engine efficiency,  and restrict our consideration to only weak coupling between the system and the bath.

Squeezing being the centerpiece in the quantum engine under study, both in the sense of changing the natural frequency of the oscillator in the isentropic phase and for the squeezed thermal bath in the isothermal phase,  we give in Sec. II  some background description of squeezed states and squeezed thermal baths.  We show how a system placed in contact with a squeezed thermal bath evolves and determine  the quantum state of the system at late times after equilibration.  This is of  interest to the problem at hand because the end state of this isothermal phase becomes the initial state of the isentropic phase.  In Sec. III  we describe the Otto cycle.  In Sec. IV we treat the simpler \textit{isentropic} phases -- no contact with a bath -- by  the Fokker-Planck-Wigner equation, where the natural frequency of the harmonic oscillator is allowed to change,  both adiabatically (gradual variation) and nonadiabatically (sudden quench). We discuss the cost versus the gain in the performance in both situations.    Finally in Sec. V we combine the results from Sec. II - IV to calculate the performance efficiency of the Otto quantum engine,  in a broader scope of  low temperature and nonadiabaticity effects. We conclude in Sec. VI.  In the Appendix we provide the coefficients and expressions which enter into the HPZ equation at late times. 

  

\section{Squeezed state for the closed/open systems}
 In this section we wish to examine more closely the nature and effects of squeezed states in the two separate phases of the Otto cycle. In the isentropic phase the oscillator is not in contact with a heat bath. Although its natural frequency is changed (sometimes the change of frequency is referred to as squeezing), it remains a closed system which follows unitary evolution.  In the isothermal phase, the system is in contact with a squeezed thermal bath and thus it constitutes an open system.  Of special interest to us is whether there exists a stationary state of the system after it interacts with a squeezed thermal bath for some time, and what state would it be after equilibration. This is needed because that state becomes the initial state of the isentropic phase in the Otto cycle.

\subsection{Brief review of squeezing}

We first  review the relevant properties of squeezing. Suppose we have a free harmonic oscillator of mass $m$ and natural frequency $\omega$, prepared in a squeezed thermal state, whose density matrix $\hat{\rho}^{(\chi)}_{\textsc{st}}$ takes the form
\begin{equation}
	\hat{\rho}^{(\chi)}_{\textsc{st}}=\hat{S}(\zeta)\hat{\rho}^{(\chi)}_{\beta}\hat{S}^{\dagger}(\zeta)\,,
\end{equation}
where $\hat{\rho}^{(\chi)}_{\beta}$ is the thermal state of the harmonic oscillator, and $\hat{S}(\zeta)$ is the squeezed operator
\begin{equation}
	\hat{S}(\zeta)=\exp\Bigl[\frac{1}{2}\,\zeta^{*}\hat{a}^{2}-\frac{1}{2}\,\zeta\hat{a}^{\dagger2}\Bigr]\,.
\end{equation}
The squeeze parameter $\zeta\in\mathbb{C}$, assumed to be a frequency-independent constant, is usually conveniently written in the polar form $\zeta=\eta\,e^{i\theta}$ with $\eta\in\mathbb{R}^{+}$ and $0\leq\theta<2\pi$. The creation and annihilation operators satisfy the standard commutation relation $[\hat{a},\hat{a}^{\dagger}]=1$, such that the displacement $\hat{\chi}$ and the corresponding conjugate momentum $\hat{p}$ of the oscillator, obeying the Heisenberg equation $\ddot{\hat{\chi}}(t)+\omega^{2}\,\hat{\chi}(t)=0$, can be expressed as
\begin{align}
	\hat{\chi}(t)&=\frac{1}{\sqrt{2m\omega}}\,\bigl(\hat{a}\,e^{-i\omega t}+\hat{a}^{\dagger}\,e^{+i\omega t}\bigr)\,,&\hat{p}(t)&=i\sqrt{\frac{m\omega}{2}}\,\bigl(\hat{a}^{\dagger}\,e^{+i\omega t}-\hat{a}\,e^{-i\omega t}\bigr)\,.
\end{align}
Here the overhead dot denotes the time derivative and $\hbar=1$ is chosen.

Since the squeezed thermal state is a Gaussian state, its statistical properties are fully described by the first two moments of $\hat{a}$, $\hat{a}^{\dagger}$. The higher moments can be obtained by the Wick expansion. Thus we need only
\begin{align}
	\langle\hat{a}\rangle_{\textsc{st}}&=0=\langle\hat{a}^{\dagger}\rangle_{\textsc{st}}\,,\notag\\
	\langle\hat{a}^{2}\rangle_{\textsc{st}}&=-e^{+i\theta}\,\sinh2\eta\,\Bigl(\langle\hat{N}\rangle_{\beta}+\frac{1}{2}\Bigr)\,,&\langle\hat{a}^{\dagger}\hat{a}\rangle_{\textsc{st}}&=\cosh2\eta\,\Bigl(\langle\hat{N}\rangle_{\beta}+\frac{1}{2}\Bigr)-\frac{1}{2}\,,\label{E:fbkdrd}
\end{align}
where the last term carries the average number of particles
\begin{equation}
	\langle\hat{N}\rangle_{\textsc{st}}=\cosh2\eta\,\langle\hat{N}\rangle_{\beta}+\sinh^{2}\eta\,,
\end{equation}
in which the factor $\langle\hat{N}\rangle_{\beta}+\frac{1}{2}$ is often conveniently written as
\begin{equation}
	\langle\hat{N}\rangle_{\beta}+\frac{1}{2}=\frac{1}{2}\,\coth\frac{\beta\omega}{2}\,.
\end{equation}
and $\beta$ is the inverse temperature of the thermal state.

The covariance matrix elements $\langle\hat{\chi}^{2}\rangle_{\textsc{st}}$, $\langle\hat{p}^{2}\rangle_{\textsc{st}}$ and $\frac{1}{2}\langle\hat{\chi}(t),\hat{p}(t)\rangle_{\textsc{st}}$ characterize the natures of the states of the oscillator. The corresponding density matrix elements can be uniquely constructed from them. For a harmonic oscillator in a  squeezed thermal state, its covariance matrix elements are related to the corresponding elements in a thermal state by
\begin{align}\label{E:deuewowo}
	\langle\hat{\chi}^{2}(t)\rangle_{\textsc{st}}&=\Bigl[\cosh2\eta-\cos(2\omega t-\theta)\,\sinh2\eta\Bigr]\,\langle\hat{\chi}^{2}\rangle_{\beta}\,,\\
	\langle\hat{p}^{2}(t)\rangle_{\textsc{st}}&=\Bigl[\cosh2\eta+\cos(2\omega t-\theta)\,\sinh2\eta\Bigr]\,\langle\hat{p}^{2}\rangle_{\beta}\,.
\end{align}
In particular the cross correlation between the canonical variables reveals information about the stationarity of the state
\begin{align}\label{E:ggbejjsd}
	\frac{1}{2}\langle\hat{\chi}(t),\hat{p}(t)\rangle_{\textsc{st}}&=\sinh2\eta\sin(2\omega t-\theta)\,\Bigl(\langle\hat{N}\rangle_{\beta}+\frac{1}{2}\Bigr)\,.
\end{align}
These elements are time dependent, in contrast to their counterparts in the thermal state. Observe that the factors before $\langle\hat{\chi}^{2}\rangle_{\beta}$ and $\langle\hat{p}^{2}\rangle_{\beta}$ are positive real numbers, so they can be smaller or larger than unity, depending on the choice of the squeeze angle $\theta$ and time $t$.  Namely, one may suppress the dispersion of the $\chi$ quadrature at the expense of the momentum uncertainty.  Thus one can use the squeeze parameter to tune the coherence of the oscillator such that, in this case, the noise level of the $\chi$ quadrature can be much lower than that of the ground state. This trick has been widely applied in quantum optics to bring down the quantum noise  in  high precision interferometer measurements such as in LIGO.

The mechanical energy of the free oscillator in a squeezed thermal state is
\begin{equation}\label{E:jderiss}
	E_{\textsc{st}}^{(\textsc{c})}=\frac{1}{2m}\langle\hat{p}^{2}\rangle_{\textsc{st}}+\frac{m\omega^{2}_{\textsc{r}}}{2}\langle\hat{\chi}^{2}\rangle_{\textsc{st}}=\bigl(\langle\hat{N}\rangle_{\textsc{st}}+\frac{1}{2}\bigr)\omega=\cosh2\eta\,E^{(\textsc{c})}_{\beta}\,,
\end{equation}
where $E^{(\textsc{c})}_{\beta}$ is the thermal energy of the oscillator, so it  does not depend on the squeeze angle $\theta$, as well as the factor $\sinh2\eta$. It is clearly seen that even if one has effectively suppressed the uncertainty in one quadrature, this induces a large uncertainty in the complementary conjugated-quadrature, such that the total energy still increases.

Finally we note the Hadamard function of the harmonic oscillator in the squeezed thermal state
\begin{align}
	G_{H,\textsc{st}}^{(\chi)}(t,t')&=\frac{1}{2}\operatorname{Tr}\Bigl[\hat{\rho}_{\textsc{st}}\bigr\{\hat{\chi}(t),\hat{\chi}(t')\bigr\}\Bigr]\notag\\
	&=\frac{1}{4m\omega}\,\coth\frac{\beta\omega}{2}\biggl\{\cosh2\eta\,\Bigl[e^{-i\omega(t-t')}+e^{+i\omega(t-t')}\Bigr]\biggr.\notag\\
	&\qquad\qquad\qquad\qquad\qquad\qquad-\biggl.\sinh2\eta\,\Bigl[e^{-i\omega(t+t')}e^{+i\theta}+e^{+i\omega(t+t')}e^{-i\theta}\Bigr]\biggr\}\,.\label{E:bgsjwer}
\end{align}
is not invariant under time translation, and thus the oscillator has nonstationary correlation, while the expressions in the brackets next to $\cosh2\eta$ is stationary. Thus it is easily seen that a squeezed thermal state is not a steady state, while a thermal state is.

\subsection{Open System: Squeezed Thermal Baths}

 We consider a harmonic oscillator of fixed natural frequency bilinearly coupled to a thermal bath, modeled by a massless quantum scalar field initially in a squeezed thermal state of temperature $\beta^{-1}$. The bath thus has a gapless spectrum. 
Once  the oscillator is brought into contact with the bath, both the oscillator and the bath will undergo nonequilibrium evolutions {even though the change of the bath in general is negligible due to its overwhelmingly larger number of degrees of freedom}. The effects of the coupling are manifested in the forms of the quantum noise from the bath and the damping force on the oscillator. The quantum noise drives the oscillator away from its initial configuration, while the damping   counteracts the influence of the driving noise. 

Compared to a thermal bath, the squeezed thermal bath has a few remarkable features. As discussed in the previous section, the quantum noise of a squeezed thermal bath is nonstationary. This immediately raises a concern whether the dissipative effect can fully counteract the driving of the nonstationary noise such that the oscillator can relax to a stationary/equilibrium state? We are    interested in the  similarities and dissimilarities in the behavior of the oscillator in such an equilibrium state compared to that  of the oscillator in the closed-system  highlighted in the previous section.

Detailed calculations presented elsewhere \cite{FDRSq}  show that when the oscillator is bilinearly coupled to the squeezed thermal bath, even though the bath gives a nonstationary driving force, the oscillator will eventually approach an equilibrium state.  Its correlation function --  the Hadamard function -- starting off nonstationary in time, will become invariant in time translation after the relaxation time. It takes the form
\begin{align}\label{E:gbdfkgjbs}
	G_{H}^{(\chi)}(t,t')=\cosh2\eta\,G_{H,\beta}^{(\chi)}(t-t')\,,
\end{align}
for $t$, $t'\gg\gamma^{-1}$ where $\gamma$ is the damping constant, and $\eta$ is the squeeze parameter of the bath in its initial configuration. The Hadamard function $G_{H,\beta}^{(\chi)}(t-t')$ gives the correlation of the oscillator when it is coupled to a plain thermal state.  In contrast to \eqref{E:bgsjwer}, we see that the nonstationary component is will be absent at late times.

Physically, when the oscillator is coupled to a bath, it behaves like a driven, damped oscillator, instead of the free oscillator in the closed system configuration. It can be shown that although the driving force coming from the bath is nonstationary, the damping will respond  in a delicate way. The strength of dissipation depends on both the dissipation kernel, determined by the form of coupling and the property of the bath field, and on the state of the oscillator's motion. Thus when the oscillator is driven by (nonstationary) quantum fluctuations of the bath, the damping will adjust itself to match the driving force in accordance, such that in the end the energy exchange between the oscillator and the bath is balanced, and the system reaches equilibration. This final state, in the weak coupling limit, is essentially a thermal state at an effective temperature $\beta^{-1}_{\textsc{s}}$, satisfying
\begin{equation}\label{E:jrdngd}
	\coth\frac{\beta_{\textsc{s}}\omega}{2}=\cosh2\eta\,\coth\frac{\beta\omega}{2}\,,
\end{equation}
in contrast to the squeezed thermal state of the free oscillator introduced in the previous section found in the literature. They have distinct features. The latter is clearly nonstationary, thus in conflict with the notion of an equilibrium state. The former is stationary, and is the very state that is often referred to at the end of the  isothermal phase of the Otto engine when its working medium is in contact with a squeezed thermal bath.  This subtle difference is not  clearly noticed in the literature, and confusion may arise when mis-identification occurs. Equally noteworthy is that this behavior, specifically,  the {\it nonstationarity of the two-point function} $G_{H}^{(\chi)}(t,t')$, will not be restored to the form in \eqref{E:bgsjwer}, that is not going back to the squeezed thermal state, even in the limit of ultraweak coupling, in stark contrast to the situation when the oscillator interacts with a plain thermal bath.  In the latter case, in the weak coupling limit, the dynamical behavior of the oscillator in the final equilibrium state is pretty much the same as that of the oscillator in the canonical thermal state in a closed system description.  {In other words, if the bath is initially in a squeezed state, the oscillator will not inherit the state of the bath at the end in the limit of vanishing coupling strength}.

In the final equilibrium state, the covariance matrix elements $\langle\hat{\chi}^{2}(\infty)\rangle$, $\langle\hat{p}^{2}(\infty)\rangle$ are given by
\begin{align}\label{E:fbkbdhkhyd}
	\langle\hat{\chi}^{2}(\infty)\rangle&=\cosh2\eta\,\langle\hat{\chi}^{2}(\infty)\rangle_{\beta}\,,&\langle\hat{p}^{2}(\infty)\rangle&=\cosh2\eta\,\langle\hat{p}^{2}(\infty)\rangle_{\beta}\,,&\frac{1}{2}\,\langle\bigl\{\hat{\chi}(\infty),\hat{p}(\infty)\bigr\}\rangle&=0\,,
\end{align}
where $\langle\hat{\chi}^{2}(\infty)\rangle_{\beta}$ and $\langle\hat{p}^{2}(\infty)\rangle_{\beta}$ represent the corresponding covariance matrix elements when the bath is initially prepared in a thermal state. The constancy of the first two expressions is distinct from their counterparts in \eqref{E:deuewowo} in the closed system description. In addition, the elements in \eqref{E:fbkbdhkhyd} do not contain the contribution from the nonstationary component, which is typically proportional to $\sinh2\eta$. In the weak coupling limit, we have
\begin{align}
	\langle\hat{\chi}^{2}(\infty)\rangle_{\beta}&=\frac{1}{2m\omega}\,\coth\frac{\beta\omega}{2}\,,&\langle\hat{p}^{2}(\infty)\rangle_{\beta}&=\frac{m\omega}{2}\,\coth\frac{\beta\omega}{2}\,,
\end{align}
so the forms of $\langle\hat{\chi}^{2}(\infty)\rangle$ and $\langle\hat{p}^{2}(\infty)\rangle$ also support the identification of the effective thermal state in \eqref{E:jrdngd}.

The internal energy, defined as the mechanical energy of the oscillator, is then given by
\begin{equation}\label{E:fkgbsd}
	E(\infty)=\frac{1}{2m}\,\langle\hat{p}^{2}(\infty)\rangle+\frac{m\omega_{\textsc{r}}^{2}}{2}\,\langle\hat{\chi}^{2}(\infty)\rangle=\cosh2\eta\times E_{\beta}(\infty)\,.
\end{equation}
The expression $E_{\beta}(\infty)$ is the internal energy of the oscillator if it is initially coupled to an unsqueezed thermal bath at temperature $\beta^{-1}$, and, in the ultra-weak coupling limit, it assumes the form
\begin{equation}
	E_{\beta}(\infty)=\frac{\omega}{2}\,\coth\frac{\beta\omega}{2}\,.
\end{equation}
The results in \eqref{E:fkgbsd} happens to take on the same functional form as the internal energy in \eqref{E:jderiss} in the closed system description, even though the squeezed thermal states in both formulations have totally different characteristics. This similitude may not come as a surprise because in the weak coupling limits, the squeeze parameter and the temperature of the system are equal to those of the bath, and because the  {constant} energy expression in \eqref{E:jderiss}   does not depend on the squeeze angle $\theta$ or the factor $\sinh 2\eta$. Therefore, even if one inadvertently mis-identified the squeezed thermal state, one would still  get the correct internal energy expression to compute the heat during the isothermal processes of the Otto engine. This happy end-result inconspicuously conceals some oversights or mis-treatments  in the literature of the nonequilibrium dynamics of quantum systems interacting with a squeezed thermal bath.

In \eqref{E:jrdngd}, since $\cosh2\eta\geq1$, the effective system temperature $\beta_{\textsc{s}}^{-1}$ is always greater than the initial bath temperature $\beta^{-1}$. It implies that if the hot thermal bath is initially squeezed, then squeezing can boost the effective temperature of the system at the end of the hot isothermal process. This can enhance the idealized Carnot efficiency. 

\subsection{Closed System:  oscillator with varying frequency \& unsqueezing}\label{S:bgkdter}
With these details available,  we now make some comments on {the equivalence between frequency modulation and squeezing, as well as the protocol of} unsqueezing.  Unsqueezing is understood as a procedure to introduce another squeezing to cancel out the preexisting squeezing in the state of the oscillator, the working medium of the Otto engine. It is often implemented during the isentropic stage that follows the hot squeezed isothermal phase in an attempt to restore the initial state in the isentropic phase back to a thermal state for the purpose of  enhancing the overall efficiency.  However, as shown in \eqref{E:gbdfkgjbs}, \eqref{E:fbkbdhkhyd} and \eqref{E:fkgbsd}, the final equilibrium state of the isothermal stage is a steady state, essentially a thermal state, stationary in time since the first two moments of the oscillator have the forms given by some canonical thermal state.  If this state serves as the initial state of the subsequent isentropic stage, then performing unsqueezing will in fact introduce nonstationarity into the oscillator. Suppose at time $t=0$, the initial time of an isentropic stage, we carry out an instantaneous unsqueezing by $\zeta^{*}=\eta^{*}\,e^{i\theta^{*}}$, then according to \eqref{E:deuewowo} and \eqref{E:ggbejjsd}, we cannot find a suitable squeeze parameter to remove the $\eta$ dependence in the covariance matrix elements in \eqref{E:fbkbdhkhyd}. 

{A variant of unsqueezing is proposed to carry out at the end of the isentropic phase on account of the possibility that the frequency modulation may  introduce squeezing. This is most easily seen in the Heisenberg picture, where the canonical operators $\hat{\chi}(t)$ and $\hat{p}(t)$ can be expressed by
\begin{align}
	\hat{\chi}(t)&=d_{1}(t)\,\hat{\chi}(0)+\frac{d_{2}(t)}{m}\,\hat{p}(0)\,,& \hat{p}(t)&=m\dot{d}_{1}(t)\hat{\chi}(0)+\dot{d}_{2}(t)\,\hat{p}(0)\,,
\end{align}
where $d_{1}(t)$ and $d_{2}(t)$ are a special set of the solution to the classical equation of motion of the parametric oscillator $\ddot{\chi}(t)+\omega^{2}(t)\,\chi(t)=0$, with the initial conditions $d_{1}(0)=1$, $\dot{d}_{1}(t)=0$ and $d_{2}(0)=0$, $\dot{d}_{2}(0)=1$. The frequency modulation, in this phase of duration $\tau$, will vary from $\omega(0)=\omega_{\textsc{h}}$ to $\omega(\tau)=\omega_{\textsc{l}}$. Then the cross correlation $\langle\{\hat{\chi}(t),\hat{p}(t)\}\rangle$ is given by
\begin{align}\label{E:fgbkdfhg}
	\frac{1}{2}\langle\bigl\{\hat{\chi}(t),\hat{p}(t)\bigr\}\rangle&=m\,d_{1}(t)\dot{d}_{1}(t)\langle\hat{\chi}^{2}(0)\rangle+\frac{1}{m}\,d_{2}(t)\dot{d}_{2}(t)\langle\hat{p}^{2}(0)\rangle\notag\\
	&\qquad\qquad\qquad\qquad\qquad+\frac{1}{2}\bigl[d_{1}(t)\dot{d}_{2}(t)+\dot{d}_{1}(t)d_{2}(t)\bigr]\langle\bigl\{\hat{\chi}(0),\hat{p}(0)\bigr\}\rangle\,,
\end{align}
where the expectation value is taken with respect to the initial state of the oscillator in the isentropic expansion phase. Since the initial state is stationary, Eq.~\eqref{E:fgbkdfhg} reduces to
\begin{align}\label{E:fgksbgksf}
	\frac{1}{2}\langle\bigl\{\hat{\chi}(\tau),\hat{p}(\tau)\bigr\}\rangle&=\frac{m}{2}\frac{d}{d\tau}\langle\hat{\chi}^{2}(\tau)\rangle=\frac{1}{2}\coth\frac{\beta_{\textsc{s}}\omega_{\textsc{h}}}{2}\,\frac{d}{d\tau}\Bigl[\frac{1}{\omega_{\textsc{h}}}\,d_{1}^{2}(\tau)+\omega_{\textsc{h}}\,d_{2}^{2}(\tau)\Bigr]\,,
\end{align}
at the end of the isentropic phase. For a harmonic oscillator of fixed frequency $\omega_{\textsc{h}}$, the expression inside the square brackets is a constant and thus \eqref{E:fgksbgksf} gives zero. However, for a general parametric oscillator, \eqref{E:fgksbgksf} is not zero, so the end state of the oscillator is not stationary and acquires squeezing. Similarly we can find $\langle\hat{\chi}^{2}(\tau)\rangle$  and $\langle\hat{p}^{2}(\tau)\rangle$ 
\begin{align}
	\langle\hat{\chi}^{2}(\tau)\rangle&=\frac{1}{2m\omega_{\textsc{h}}}\,\coth\frac{\beta_{\textsc{s}}\omega_{\textsc{h}}}{2}\,\Bigl[d_{1}^{2}(\tau)+\omega_{\textsc{h}}^{2}d_{2}^{2}(\tau)\Bigr]\,,\\
	\langle\hat{p}^{2}(\tau)\rangle&=\frac{m\omega_{\textsc{h}}}{2}\,\coth\frac{\beta_{\textsc{s}}\omega_{\textsc{h}}}{2}\,\Bigl[\frac{1}{\omega_{\textsc{h}}^{2}}\,\dot{d}_{1}^{2}(\tau)+\dot{d}_{2}^{2}(\tau)\Bigr]\,,\label{E:rgugbr}
\end{align}
for this end state. Then according to (3.9)--(3.11) in~\cite{NEqFE}, for the oscillator with  frequency $\omega$ in  a general squeezed thermal state, parametrized by the squeezed parameters $(\eta_{\textsc{s}},\psi_{\textsc{s}})$ and an inverse temperature-like parameter $\vartheta$, the covariance matrix elements take on the form
\begin{align}
	\langle\hat{\chi}^{2}\rangle&=\frac{1}{2m\omega_{\textsc{l}}}\,\coth\frac{\vartheta}{2}\,\bigl(\cosh2\eta_{\textsc{s}}-\sinh2\eta_{\textsc{s}}\,\cos\psi_{\textsc{s}}\bigr)\,,\label{E:fkgdbk1}\\
	\langle\hat{p}^{2}\rangle&=\frac{m\omega_{\textsc{l}}}{2}\,\coth\frac{\vartheta}{2}\,\bigl(\cosh2\eta_{\textsc{s}}+\sinh2\eta_{\textsc{s}}\,\cos\psi_{\textsc{s}}\bigr)\,,\\
	\frac{1}{2}\langle\bigl\{\hat{\chi},\hat{p}\bigr\}\rangle&=-\frac{1}{2}\,\coth\frac{\vartheta}{2}\,\sinh2\eta_{\textsc{s}}\,\sin\psi_{\textsc{s}}\,.\label{E:fkgdbk3}
\end{align}
Comparing \eqref{E:fkgdbk1}--\eqref{E:fkgdbk3} with \eqref{E:fgksbgksf}--\eqref{E:rgugbr}, we identify $\vartheta=\beta_{\textsc{s}}\omega_{\textsc{h}}$, and the effective squeeze  parameter $\eta_{\textsc{s}}$
\begin{align}\label{E:fgbkjsbgdf}
	\eta_{\textsc{s}}=\frac{1}{2}\cosh^{-1}\biggl\{\frac{1}{2}\Bigl[\frac{\omega_{\textsc{l}}}{\omega_{\textsc{h}}}\,d_{1}^{2}(\tau)+\omega_{\textsc{h}}\omega_{\textsc{l}}\,d_{2}^{2}(\tau)+\frac{1}{\omega_{\textsc{h}}\omega_{\textsc{l}}}\,\dot{d}_{1}^{2}(\tau)+\frac{\omega_{\textsc{h}}}{\omega_{\textsc{l}}}\,\dot{d}_{2}^{2}(\tau)\Bigr]\biggr\}\,,
\end{align}
of the oscillator at the end of the isentropic phase. We use \eqref{E:fkgdbk3} to find the corresponding squeeze angle $\psi_{\textsc{s}}$. These can then be  used to unsqueeze the oscillator state.}  {The expression in the curly brackets can serve as a measure of the degree of nonadiabaticity, which is essentially the ratio of the energy at the end of an arbitrary parametric process to its counterpart, the energy for the adiabatic process if the oscillator is initially in the ground state. 

{We are led to an interesting competition during the non-adiabatic modulation. On one hand, the non-adiabatic frequency change of the parametric oscillator degrades the output work due to unwanted excitations. On the other hand, allowing for unsqueezing, as shown later, can  enhance the efficiency at the expense of the agent that performs the action. This will be explored in greater detail in Sec.~VI.}

\section{Otto cycle}

We have highlighted the relevant theories for closed and open systems behind the isentropic and the isothermal phases of an Otto cycle.   In this section we briefly describe the operational protocols of the Otto cycle before entering into the details towards the assessment of engine efficiency in the following two sections.   
The Otto cycle consists of the following four stages in two alternating phases (for illustration we use the same labeling as in  
Fig. 1b of ~\cite{Manzano}): 
\begin{enumerate}[1)]
	\item isentropic compression phase ($A\mapsto B$): At point $A$,  the oscillator is isolated from a thermal bath, its initial state is a thermal state of temperature $\beta_{\textsc{l}}^{-1}$ . The natural frequency of the oscillator is then raised to $\omega_{\textsc{h}}$ from $\omega_{\textsc{l}}<\omega_{\textsc{h}}$ by an external agent. In this phase, the oscillator evolves as a closed system, so there is no heat exchange, and the work done by the external agent will solely change the system's internal energy. For our theoretical inquiry we do not limit the isentropic process to be adiabatically slow even though the conventional wisdom tells that the output work from the engine tends to be optimal for an adiabatically slow modulation of the frequency because the excitation to higher energy levels is less likely to occur. {Unsqueezing may be introduced at the end of the isentropic phase to reverse the squeezing brought about by the nonadiabatic frequency modualtion.}
	\item hot isothermal phase ($B\mapsto C$): At point $B$, the oscillator with frequency $\omega_{\textsc{h}}$ is brought back into contact with a hot squeezed thermal bath at temperature $\beta_{\textsc{h}}^{-1}$. The oscillator will undergo nonequilibrium relaxation to an equilibrium state and end up at point $C$ of the cycle. In this phase the energy exchange with the bath, identified as heat, will change the internal energy of the oscillator.
	\item isentropic expansion phase ($C\mapsto D$): Here the oscillator frequency is lowered down to $\omega_{\textsc{l}}$. Since it is isolated from the thermal bath, the oscillator only exchanges work with the outside agent that modulates the frequency.   {In contrast to the isentropic phase $(A\mapsto B)$, we note that in the previous isothermal phase ($B\mapsto C$), the bath is squeezed at the outset, so in principle an unsqueezing is needed at the beginning of the current phase to remove the squeezing of the system during the isothermal phase. However, from our earlier discussions,  if the oscillator in the isothermal phase prior to this isentropic phase had undergone full relaxation, then it is not necessary to do so. We only need to carry out unsqueezing at the end of the isothermal phase if the expansion there is non-adiabatic.}
	\item cold isothermal phase ($D\mapsto A$):  At point $D$, the  oscillator of frequency $\omega_{\textsc{l}}$ is placed in contact with a cold thermal bath at temperature $\beta_{\textsc{l}}^{-1}$. The oscillator is then relaxed to its equilibrium state at point $C$, while heat is exchanged between the oscillator and the bath. In the weak coupling limit, this final state will be a thermal state, not a squeezed thermal state. This completes the cycle. 
\end{enumerate}

\section{Isentropic stages: Frequency change in a closed system}\label{isent}

In this section, we look into the oscillator dynamics during the isentropic phases, that is, stage 1) and 3) of the Otto cycle defined in the last section. The oscillator is completely isolated from the thermal bath,  but  driven by an external agent that modulates  its  natural frequency $\omega(t)$.  Since there is no bath, we consider the parametric oscillator as a closed system. Assuming the oscillator has a mass $m$, and its canonical variables $\chi$, the displacement, and the conjugate momentum $p$ satisfy the commutation relation $[\hat{\chi},\hat{p}]=i$. The Hamiltonian of the oscillator then takes the form
\begin{equation}
	\hat{H}_{\textsc{s}}(t)= \frac{\hat{p}^2}{2m} + \frac{m\omega^2(t)}{2}\,\hat{\chi}^2\,,
\end{equation}
for $0\leq t\leq\tau$, where $\tau$ is the duration of the process. For example, in the isentropic phase $A$) the oscillator frequency changes from $\omega(0)=\omega_{\textsc{l}}$ to $\omega(\tau)=\omega_{\textsc{h}}$. For simplicity,  to capture the essential physics, we assume $\omega^{2}(t)$ monotonically increases during this stage according to
\begin{align}\label{E:fbksfg}
	\omega^{2}(t)&=\omega^{2}_{\textsc{l}}+\frac{t}{\tau}\bigl(\omega^{2}_{\textsc{h}}-\omega^{2}_{\textsc{l}}\bigr)\,,&0\leq t\leq\tau\,.
\end{align}
The time scale $\tau$ describes how fast the oscillator's natural frequency varies and thus is a measure of the (non)adiabaticity of this process, that is, $\dot{\omega}/\omega^{2}$. A small $\tau$ depicts the qualitative features of sudden quenching while the a large value of $\tau$ can approch ideal adiabaticity. We will not a priori assume that the frequency modulation is adiabatically slow, but for a typical Otto engine to produce a positive net work output,   
$\omega_{\textsc{h}} > \omega_{\textsc{c}}$ is usually required.

\begin{figure}
 	\centering
 	\scalebox{0.7}{\includegraphics{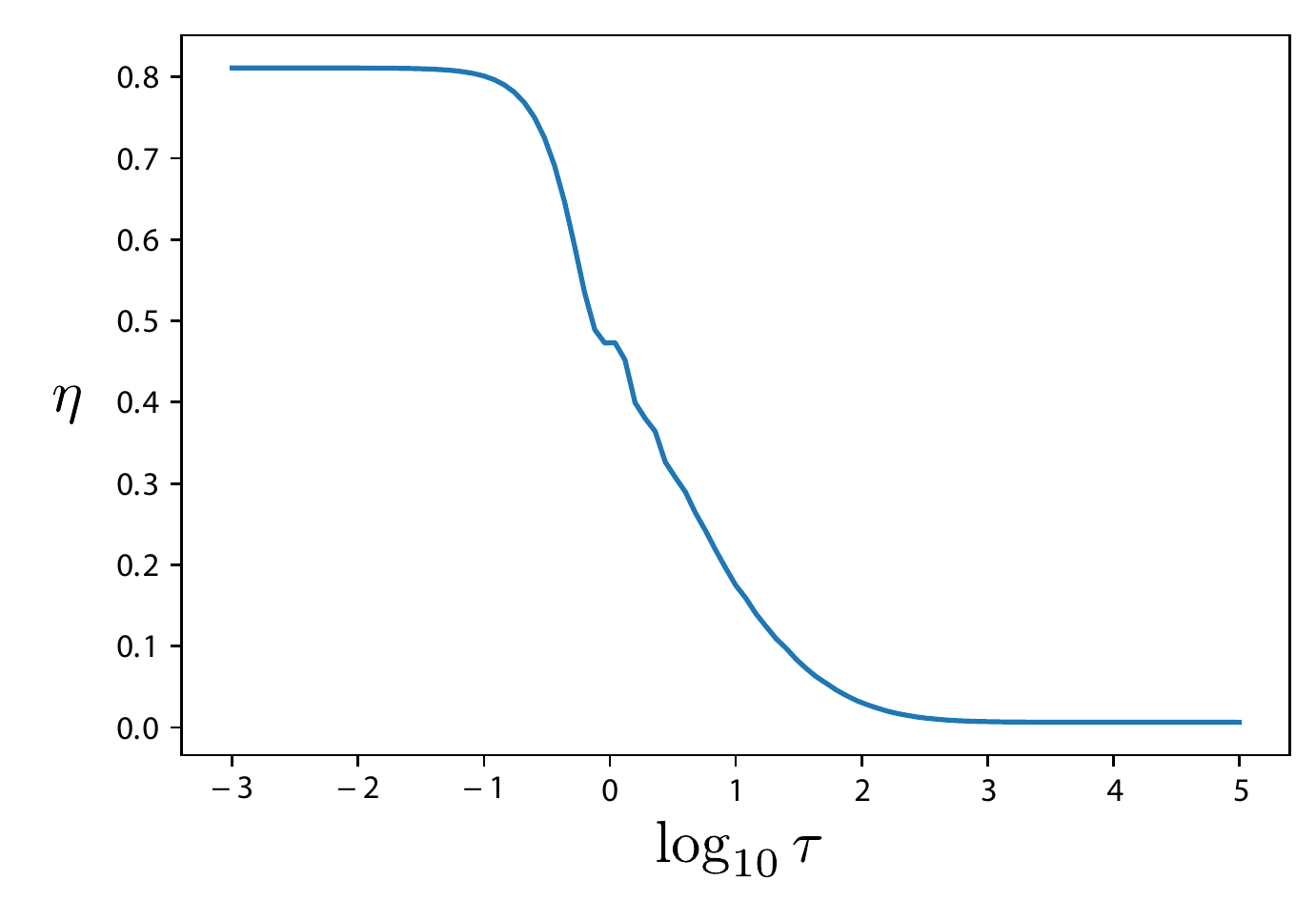}}
 	\caption{Here we show how the squeezing of the oscillator at the end of the isentropic stage depends on the speed of its frequency modulation. The parameter $\tau$ is the duration of the frequency modulation. The initial state of the isentropic phase is given by the final equilibrium state of the oscillator in the previous isothermal phase where it is in contact with an Ohmic thermal bath of temperature $\beta_{\textsc{l}}^{-1}= 0.001$. We choose $\omega_{\textsc{l}} = 1$, $\omega_{\textsc{h}} = 5\omega_{\textsc{l}}$, $m=1$, $\gamma=2\times 10^{-5}$, and the cutoff for the bath $\Lambda_{\textsc{l}}=1000$.}\label{fig:r-driving}
 \end{figure}

We further assume that the initial state of this phase is a generic Gaussian state. Since the working medium of the engine is modeled by a parametric harmonic  oscillator throughout this stage, it is guaranteed that the end state is another Gaussian state. Therefore, the Wigner function of the system during the isentropic stage takes the generic form 
\begin{equation}
	\mathcal{W}(\chi,p,t) = N\, \exp\Bigl[\mathcal{A}(t)\,\chi^2 + \mathcal{B}(t)\,p^2 + \mathcal{C}(t)\,\chi p\Bigr]\,,\label{gauss}
\end{equation}
where the normalization factor $N$ is time-independent while the coefficients $\mathcal{A}$, $\mathcal{B}$ and $\mathcal{C}$ in the exponent are time-dependent. The Wigner function satisfies the Wigner-Fokker-Planck equation, e.g.~\cite{HalYu96}
\begin{equation}
	\frac{\partial \mathcal{W}}{\partial t} = -\frac{p}{m} \frac{\partial \mathcal{W}}{\partial \chi} + m\omega^2(t)\,\chi\,\frac{\partial \mathcal{W}}{\partial p}\,, \label{wfp}
\end{equation}
so its solution dictates the time evolution of the parametric oscillator. Plugging in Eq.~\eqref{gauss} to Eq.~\eqref{wfp} gives a coupled set of differential equations for the coefficients $\mathcal{A}$, $\mathcal{B}$ and $\mathcal{C}$,
\begin{align}\label{E:fgbksdgs}
\frac{d}{dt}\mathcal{A}(t)&=m\omega^2(t)\,\mathcal{C}(t)\,,&\frac{d}{dt}\mathcal{B}(t) &=-\frac{1}{m}\,\mathcal{C}(t) \,,&\frac{d}{dt}\mathcal{C}(t) &= 2\Bigl[m\omega^2(t)\,\mathcal{B}(t)-\frac{1}{M}\,\mathcal{A}(t)\Bigr]\,,
\end{align}
with the initial conditions of $\mathcal{A}$, $\mathcal{B}$ and $\mathcal{C}$ determined by the initial state of the isentropic phase. Given an arbitrary function $\omega (t)$, it is not easy to find an analytical, closed-form solution for this set of differential equations barring a few exceptions such as \eqref{E:fbksfg}. A rather general analytical treatment is developed in~\cite{HK53}, where the unitary time evolution operator in the $\chi$ representation can be constructed by a special set of fundamental solutions to the equation of motion of the parametric oscillator. Here we resort to numerical methods to get the solutions of \eqref{E:fgbksdgs}. For an arbitrary speed of frequency modualtion, Fig~\ref{fig:r-driving}, show the variation of the squeezing at the end of frequency modulation with the transition time $\tau$. The sudden change limit results in a squeeze parameter $\eta$ given by $\frac{1}{2}\ln\omega_{\textsc{h}}/\omega_{\textsc{l}}$. On the other hand, the adiabatic limit shows that the squeezing parameter remains the same after the driving. The intermediate regime shows a monotonous behavior between these two regimes.

Since the oscillator is isolated from the bath, the external agent that modifies the oscillator frequency will have energy exchange with the oscillator in the form of work. By energy conservation the work is related to the change of the internal energy of the oscillator. During the isentropic phase $A\mapsto B$, that is, $\omega_{\textsc{l}}\mapsto\omega_{\textsc{h}}$, the work is given by the internal energy difference
\begin{equation}
	W_{\textsc{bc}}=E_{\textsc{b}}-E_{\textsc{a}}\,,
\end{equation}
where $E_{\textsc{b}}$ for example represents the internal energy of the oscillator at point $B$. The internal energy is the expectation value of the Hamiltonian $\hat{H}_{\textsc{s}_{i}}$ for $i=\textsc{l}$, $\textsc{h}$,
\begin{equation}
	\hat{H}_{\textsc{s}_{i}}=\frac{\hat{p}^{2}}{2m}+\frac{m\omega_{i}^{2}}{2}\hat{\chi}^{2}\,.
\end{equation}
We adopt the convention that when $W$ is positive, work is done to the oscillator, but when $W<0$, the oscillator outputs work. For weak oscillator-bath coupling we immediately have $E_{\textsc{a}}$ given by
\begin{equation}
	E_{\textsc{a}}=\frac{\omega_{\textsc{l}}}{2}\coth\frac{\beta_{\textsc{l}}\omega_{\textsc{l}}}{2}\,.
\end{equation}
In the graduate change {(AD)} limit, we have
\begin{equation}\label{E:nderds}
	W^{(\textsc{ad})}_{\textsc{ab}}=\Bigl(\frac{\omega_{\textsc{h}}}{2}-\frac{\omega_{\textsc{l}}}{2}\Bigr)\coth\frac{\beta_{\textsc{l}}\omega_{\textsc{l}}}{2}>0\,,
\end{equation}
since the level population does not change with time. On the other hand,  in the sudden change {(SC)} limit, the work is given by
\begin{equation}\label{E:kfgbkdjsd}
	W^{(\textsc{sc})}_{\textsc{ab}}=\frac{\omega_{\textsc{h}}^{2}-\omega_{\textsc{l}}^{2}}{4\omega_{\textsc{l}}}\,\coth\frac{\beta_{\textsc{l}}\omega_{\textsc{l}}}{2}>0\,,
\end{equation}
because the state cannot adapt itself fast enough to match the configuration change. We see that typically if we require $\omega_{\textsc{h}}>\omega_{\textsc{l}}$, then the external agent needs to input more energy to the oscillator in order to arrive at the same frequency change. This is a consequence of the fact that when the parametric process is non-adiabatic, additional energy goes into (is `wasted')  exciting the oscillator.

Similarly for the other isentropic phase $C\mapsto D$, i.e., $\omega_{\textsc{h}}\mapsto\omega_{\textsc{l}}$, the internal energy of the oscillator at $C$ is given by
\begin{equation}\label{E:gkssds}
	E_{\textsc{c}}=\cosh2\eta\times\frac{\omega_{\textsc{h}}}{2}\coth\frac{\beta_{\textsc{h}}\omega_{\textsc{h}}}{2}=\frac{\omega_{\textsc{h}}}{2}\coth\frac{\beta_{\textsc{s}}\omega_{\textsc{h}}}{2}\,,
\end{equation}
because the hot thermal bath is initially prepared in a squeezed thermal state. After the frequency modulation, the work during this phase is given by
\begin{equation}
	W_{\textsc{cd}}=E_{\textsc{d}}-E_{\textsc{c}}\,.
\end{equation}
Assuming no additional unsqueezing, the work would depend  on the speed of frequency modulation.  In the adiabatic limit  $W_{\textsc{cd}}$ is given by
\begin{equation}\label{E:bgfket}
	W^{(\textsc{ad})}_{\textsc{cd}}=\cosh2\eta\,\Bigl(\frac{\omega_{\textsc{l}}}{2}-\frac{\omega_{\textsc{h}}}{2}\Bigr)\coth\frac{\beta_{\textsc{h}}\omega_{\textsc{h}}}{2}<0\,,
\end{equation}
but in the sudden limit
\begin{equation}\label{E:gbhksgf}
	W^{(\textsc{sc})}_{\textsc{cd}}=\cosh2\eta\,\frac{\omega_{\textsc{l}}^{2}-\omega_{\textsc{h}}^{2}}{4\omega_{\textsc{h}}}\,\coth\frac{\beta_{\textsc{h}}\omega_{\textsc{h}}}{2}<0\,.
\end{equation}
Clearly the output work is greater in the adiabatic limit when $\omega_{\textsc{h}}>\omega_{\textsc{l}}$. {For an arbitrary speed of frequency modulation, we can write the internal energy at the end of the parametric process formally as
\begin{equation}
	E_{\textsc{d}}=\cosh\eta_{\textsc{s}}\times\frac{\omega_{\textsc{h}}}{2}\,\coth\frac{\beta_{\textsc{s}}\omega_{\textsc{h}}}{2}
\end{equation}
where the factor $\cosh\eta_{\textsc{s}}$ generically accounts for the squeezing due to frequency modulation, and $\eta_{\textsc{s}}$ is given by \eqref{E:fgbkjsbgdf}. Note that this is a different squeeze parameter from the one $\eta$  used for the squeezed bath. If we introduce unsqueezing at the end of the isentropic phase, then we restore, apart from a rotation, the state of the oscillator to the one according to the adiabatic modulation, thus recovering the result given by \eqref{E:bgfket} for any modulation speed. This can be viewed as some type of  `shortcut to adiabaticity' (STA) implemented in the isentropic phase~\cite{STA}. The implementation is not limited to the specific form of frequency modulation \eqref{E:fbksfg}. From the discussion in Sec.~\ref{S:bgkdter}, we see it is quite general. We can always identify the suitable $\zeta_{\textsc{s}}$, in terms of the fundamental solutions $d_{1,2}(\tau)$, to perform unsqueezing given any functional form of $\omega(t)$.

From Eq.~\eqref{E:nderds} and Eq.~\eqref{E:bgfket}, we learn that in the adiabatic limit, the external agent pumps less energy to the oscillator during the isentropic compression phase $AB$, but the oscillator outputs more work to the outside in the isentropic expansion phase $CD$. The net work in the adiabatic limit is thus
\begin{equation}
	W^{(\textsc{ad})}_{\textsc{tot}}=W^{(\textsc{ad})}_{\textsc{ab}}+W^{(\textsc{ad})}_{\textsc{cd}}=-\Bigl(\frac{\omega_{\textsc{h}}}{2}-\frac{\omega_{\textsc{l}}}{2}\Bigr)\Bigl(\cosh2\eta\,\coth\frac{\beta_{\textsc{h}}\omega_{\textsc{h}}}{2}-\coth\frac{\beta_{\textsc{l}}\omega_{\textsc{l}}}{2}\Bigr)\,.
\end{equation}
We see that the magnitude of the total work depends on the combinations of $\eta$, $\beta_{\textsc{h}}\omega_{\textsc{h}}$ and $\beta_{\textsc{l}}\omega_{\textsc{l}}$ with the constraints $\eta\geq0$, $\beta_{\textsc{h}}<\beta_{\textsc{l}}$ and $\omega_{\textsc{h}}>\omega_{\textsc{l}}$, so the design of the engine is to find an optimal combination to have the maximal output work for a given incoming heat, that is, $W_{\textsc{tot}}<0$ and maximal $\lvert W_{\textsc{tot}}\rvert$. At least when $\beta_{\textsc{h}}\omega_{\textsc{h}}<\beta_{\textsc{l}}\omega_{\textsc{l}}$, the magnitude of the total work increases with growing $\eta$, but independent of $\theta$. Furthermore, since $\cosh2\eta\geq1$, we may not want to put additional squeezing in the cold thermal bath if we want to extract maximal total work. Meanwhile the condition $\beta_{\textsc{h}}\omega_{\textsc{h}}<\beta_{\textsc{l}}\omega_{\textsc{l}}$ implies
\begin{equation}
	\frac{\omega_{\textsc{l}}}{\omega_{\textsc{h}}}>\frac{\beta_{\textsc{h}}}{\beta_{\textsc{l}}}\,,
\end{equation}
and in turn leads to the well known fact that the efficiency of the Otto engine is always less than that of the Carnot engine.

The $\theta$-independence implies that the result is independent of which quadrature is squeezed. This is very different from the application of squeezing in LIGO, where a particular quadrature has to be squeezed such that the effects of the corresponding quantum noise are suppressed in order to maintain high coherence. In the context of heat engine, the independence of $\theta$ can be traced to the internal energy of the oscillator at the end of the hot isothermal phase \eqref{E:gkssds}, which is also independent of $\theta$. Physically speaking, when we have suppressed the uncertainty in one quadrature, a large uncertainty in the complementary conjugated quadrature will be induced. Thus, if the elastic energy of the oscillator is suppressed by a factor $e^{-2\eta}$, the kinetic energy is enhanced by $e^{+2\eta}$ such that the total energy still gets bigger by $e^{+2\eta'}\sim\cosh2\eta'$, and vice versa.

In the quench limit, Eqs.~\eqref{E:kfgbkdjsd} and \eqref{E:gbhksgf} provide the total work
\begin{equation}
	W^{(\textsc{sc})}_{\textsc{tot}}=W^{(\textsc{sc})}_{\textsc{ab}}+W^{(\textsc{sc})}_{\textsc{cd}}=-\frac{1}{4}\Bigl(\omega_{\textsc{h}}^{2}-\omega_{\textsc{l}}^{2}\Bigr)\Bigl(\frac{1}{\omega_{\textsc{h}}}\,\cosh2\eta\,\coth\frac{\beta_{\textsc{h}}\omega_{\textsc{h}}}{2}-\frac{1}{\omega_{\textsc{l}}}\,\coth\frac{\beta_{\textsc{l}}\omega_{\textsc{l}}}{2}\Bigr)\,,
\end{equation}
or, more compactly,
\begin{equation}\label{E:fmsgsmfg}
	W^{(\textsc{sc})}_{\textsc{tot}}=-\frac{1}{4}\Bigl(\omega_{\textsc{h}}^{2}-\omega_{\textsc{l}}^{2}\Bigr)\Bigl(\frac{1}{\omega_{\textsc{h}}}\,\coth\frac{\beta_{\textsc{s},\textsc{h}}\omega_{\textsc{h}}}{2}-\frac{1}{\omega_{\textsc{l}}}\,\coth\frac{\beta_{\textsc{s},\textsc{l}}\omega_{\textsc{l}}}{2}\Bigr)\,,
\end{equation}
where $\beta_{\textsc{s},\textsc{h}}$ is introduced according to \eqref{E:jrdngd} and $\beta_{\textsc{s},\textsc{l}}=\beta_{\textsc{l}}$ because there is no squeezing in the low-temperature thermal bath. Here we need a stronger condition 
\begin{equation}
	\frac{1}{\omega_{\textsc{h}}}\,\coth\frac{\beta_{\textsc{s},\textsc{h}}\omega_{\textsc{h}}}{2}>\frac{1}{\omega_{\textsc{l}}}\,\coth\frac{\beta_{\textsc{s},\textsc{l}}\omega_{\textsc{l}}}{2}
\end{equation}
to guarantee an output work, rather than $\beta_{\textsc{h}}\omega_{\textsc{h}}<\beta_{\textsc{l}}\omega_{\textsc{l}}$. In addition we can verify that the magnitude of the output work in the sudden quench case is always smaller than that in the adiabatic limit. Finally, squeezing has the same effect on the output work as in the adiabatic process.

Next we move on to the isothermal phases.

\section{Isothermal stages:  Squeezed thermal baths}\label{eq}

In an isothermal stage our system of one quantum harmonic oscillator with a fixed frequency is  brought into contact with a squeezed thermal bath.  We shall allow the system to evolve to relaxation whose time scale is determined by the inverse damping constant $\gamma^{-1}$.  The existence of a steady state wherein the energy exchange between the oscillator and the bath is in balance and the correlation function of the oscillator becomes stationary, that is, invariant in time translation, is demonstrated in \cite{FDRSq}.  We rely on the HPZ equation for a Brownian oscillator in a squeezed thermal bath derived in~\cite{HM94} and provide the  late time behavior of its coefficient functions in the Appendix.  We perform numerical computations from these expressions to obtain the quantities we need for the isothermal phases of the Otto engine.

\begin{figure}
	\centering
  	\scalebox{0.4}{\includegraphics{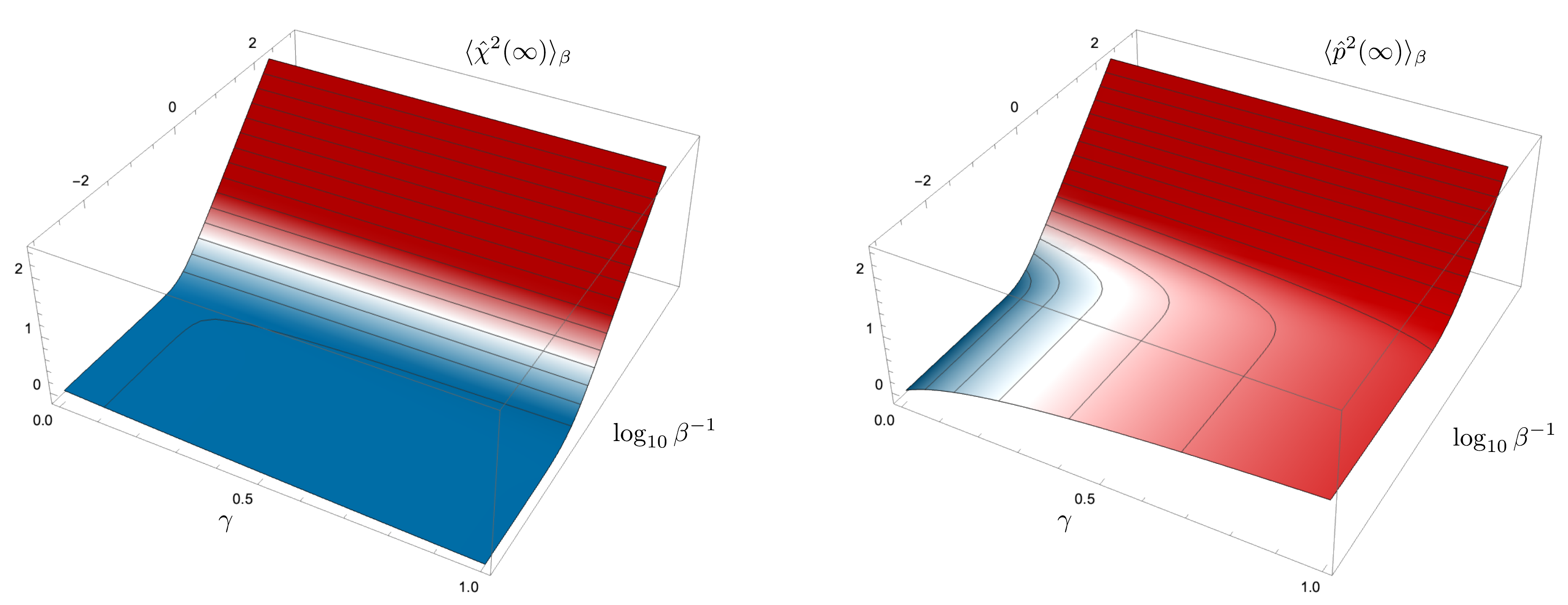}}
	\caption{Dependences of the late-time covariance matrix elements $\langle\hat{\chi}^{2}(\infty)\rangle_{\beta}$ and $\langle\hat{p}^{2}(\infty)\rangle_{\beta}$ on the damping constant $\gamma$ and the initial bath temperature $\beta^{-1}$ indicating the stark differences between  strong and weak coupling, low and high temperature. The former is scaled with respect to $(2m\omega)^{-1}$ and the latter to $m\omega/2$. The variables $\gamma$ and $\beta^{-1}$ are in the unit of $\omega$. Here we choose $m=1\times\omega$ and the cutoff scale $\Lambda=1000\omega$.}
\label{fig:Cov}
\end{figure}

With the Hamiltonian for an oscillator of mass $m$ and renormalized frequency $\omega$ given by
\begin{equation}
	\hat{H}_{\textsc{r}} =\frac{\hat{p}^2}{2m}+\frac{m\omega^2}{2}\,\hat{\chi}^2\,,
\end{equation}
the Wigner function after the system reaches a steady state takes the generic form
\begin{equation}\label{wsees}
	\mathcal{W}_{\textsc{ss}}(\chi,p)=\frac{1}{4\pi\sqrt{ab}}\,\exp\Bigl[\frac{p^2}{4a}+\frac{\chi^2}{4b}\Bigr]\,, 
\end{equation}
where the coefficients $a$, $b$ are related to the covariance matrix elements $\langle\hat{\chi}^{2}(\infty)\rangle$, $\langle\hat{p}^{2}(\infty)\rangle$ by
\begin{align}
	\langle\hat{\chi}^{2}(\infty)\rangle&=-2b=\cosh2\eta\,\frac{2\gamma}{m\pi} \int_0^{\Lambda}\!d\kappa\; \kappa\,\coth\frac{\beta\kappa}{2}\,\frac{1}{\lvert(\gamma+i\kappa)^2 +\omega_{\text{r}}^2\rvert^2}\,,\label{E:beire1}\\
	\langle\hat{p}^{2}(\infty)\rangle&=-2a=\cosh2\eta\,\frac{2m\gamma}{\pi} \int_0^{\Lambda}\!d\kappa\; \kappa^{3}\coth\frac{\beta\kappa}{2}\,\frac{1}{\lvert(\gamma+i\kappa)^2 +\omega_{\text{r}}^2\rvert^2}\,,\label{E:beire2}\\
	\langle\bigl\{\hat{\chi}(\infty),\hat{p}(\infty)\bigr\}\rangle&=0\,,
\end{align}
if the bath is initially in a squeezed thermal state at temperature $\beta^{-1}$ and squeeze parameter $\zeta=\eta\,e^{i\theta}$. Here $\infty$ in the argument of the covariance matrix elements means that these elements are evaluated at the final equilibrium state. The frequency $\omega_{\text{r}}$ denotes the resonance frequency, defined by $\omega^{2}_{\text{r}}=\omega^{2}-\gamma^{2}$. In the weak oscillator-bath coupling limit  $\omega$ and $\omega_{\text{r}}$ are approximately the same. The integral expression in $\langle\hat{p}^{2}(\infty)\rangle$ is not well defined so {we introduce a frequency cutoff $\Lambda$ in the upper limit of the $\kappa$-integral} to suppress the excessive high frequency contributions. {The cutoff parameter is typically the highest energy scale below which} the theory remains valid. {In~Fig.~\ref{fig:Cov}, we show the typical dependences of the covariance matrix elements $\langle\hat{\chi}^{2}(\infty)\rangle_{\beta}$ and $\langle\hat{p}^{2}(\infty)\rangle_{\beta}$ at late times when the oscillator is in a plain thermal bath of temperature $\beta^{-1}$. They differ from $\langle\hat{\chi}^{2}(\infty)\rangle$ and $\langle\hat{p}^{2}(\infty)\rangle$ by a constant factor $\cosh2\eta$. We mention several salient features:  1) When the bath temperature is sufficiently high, both covariance matrix elements take on the classical values, which have very weak dependence on $\gamma$. It reflects that the quantum nature of the oscillator is heavily suppressed by large thermal fluctuations of the bath. 2) At low temperatures, they deviate from their counterparts in a closed system, with a larger damping constant $\gamma$. 3) Since they are rescaled with respect to their closed-system counterparts, the state of the oscillator is slightly squeezed~\cite{NEqFE} (due to finite oscillator-bath coupling there unlike the squeezing due to frequency modulation or a fixed squeezed thermal bath here). 4) $\langle\hat{p}^{2}(\infty)\rangle_{\beta}$ has a stronger dependence on $\gamma$ at low temperatures because typically it contains a cutoff-dependent term like $\gamma\ln(\Lambda/\omega)$, aimed at the ultraviolate behavior of the vacuum fluctuations of the bath field.

We note that both elements in \eqref{E:beire1} and \eqref{E:beire2} are proportional to the factor $\cosh2\eta$, independent of the squeeze angle $\theta$, so the initial phase information of the bath is lost. The vanishing of the cross correlation between the canonical variable operators also reflects the stationarity of the final state. These properties show that the final steady state is a thermal state at a temperature $\beta_{\textsc{s}}^{-1}$ different from the bath's initial temperature $\beta^{-1}$. Essentially in the weak coupling limit, we find 
\begin{equation}
	\coth\frac{\beta_{\textsc{s}}\omega}{2}=\cosh2\eta\,\coth\frac{\beta\omega}{2}\,.
\end{equation}
Since $\cosh2\eta\geq1$, we always have $\beta_{\textsc{s}}\leq\beta$ due to the fact that $\coth z$ is a monotonically decreasing function of $z$. This implies $\beta_{\textsc{s}}^{-1}\geq\beta^{-1}$, and the squeezing always effectively raises the oscillator temperature, independent of the squeeze angle $\theta$. For a given initial state of the oscillator, more heat from the hot thermal bath goes into the oscillator with a larger squeeze parameter $\eta$, following \eqref{E:gkssds} and the argument in the previous section. This feature can be useful in boosting the theoretical Carnot efficiency for a fixed cold thermal bath temperature.

Since there is no external agent, the heat exchange with the bath is what changes the internal energy of the oscillator.  The amount of heat injected into the oscillator or released to the bath is given by the difference of the internal energy of the oscillator between its initial and final state of the isothermal process. For the hot isothermal phase ($B\mapsto C$), in the weak coupling limit, we already know the internal energy of the oscillator at $C$ is given by \eqref{E:gkssds}. Thus the heat exchange in this phase is
\begin{equation}
	Q_{\textsc{in}}=E_{\textsc{c}}-E_{\textsc{b}}\,,
\end{equation}
where $B$ denotes the final state of the previous isentropic phase. The positive value of heat means that the heat flows into the oscillator. If the isentropic phase ($A\mapsto B$) are adiabatically slow, then $E_{\textsc{b}}$ is given by
\begin{equation}
	E_{\textsc{b}}=\frac{\omega_{\textsc{h}}}{2}\,\coth\frac{\beta_{\textsc{l}}\omega_{\textsc{l}}}{2}\,,
\end{equation}
which gives the corresponding injected heat
\begin{equation}
	Q_{\textsc{in}}=E_{\textsc{c}}-E_{\textsc{b}}=\frac{\omega_{\textsc{h}}}{2}\Bigl(\cosh2\eta\,\coth\frac{\beta_{\textsc{h}}\omega_{\textsc{h}}}{2}-\coth\frac{\beta_{\textsc{l}}\omega_{\textsc{l}}}{2}\Bigr)\,.
\end{equation}
Thus in the adiabatic limit, the efficiency $\xi^{(\textsc{ad})}$ of the Otto engine in contact with the hot squeezed bath is given by
\begin{equation}
	\xi^{(\textsc{ad})}=\frac{\lvert W^{(\textsc{ad})}_{\textsc{tot}}\rvert}{Q_{\textsc{in}}}=\frac{\Bigl(\dfrac{\omega_{\textsc{h}}}{2}-\dfrac{\omega_{\textsc{l}}}{2}\Bigr)\Bigl(\cosh2\eta\,\coth\dfrac{\beta_{\textsc{h}}\omega_{\textsc{h}}}{2}-\coth\dfrac{\beta_{\textsc{l}}\omega_{\textsc{l}}}{2}\Bigr)}{\dfrac{\omega_{\textsc{h}}}{2}\Bigl(\cosh2\eta\,\coth\dfrac{\beta_{\textsc{h}}\omega_{\textsc{h}}}{2}-\coth\dfrac{\beta_{\textsc{l}}\omega_{\textsc{l}}}{2}\Bigr)}=1-\frac{\omega_{\textsc{l}}}{\omega_{\textsc{h}}}\,.
\end{equation}
The is the optimal efficiency of the Otto engine, and it is independent of the squeezing in the hot thermal bath and independent of the temperatures of the baths.

In comparison, if both isentropic phases undergo a sudden frequency change, then the efficiency $\xi^{(\textsc{sc})}$ of the engine is given by
\begin{equation}
	\xi^{(\textsc{sc})}=\frac{\lvert W^{(\textsc{sc})}_{\textsc{tot}}\rvert}{Q_{\textsc{in}}}=\frac{\frac{1}{4}\Bigl(\omega_{\textsc{h}}^{2}-\omega_{\textsc{l}}^{2}\Bigr)\Bigl(\dfrac{1}{\omega_{\textsc{h}}}\,\coth\dfrac{\beta_{\textsc{s},\textsc{h}}\omega_{\textsc{h}}}{2}-\dfrac{1}{\omega_{\textsc{l}}}\,\coth\dfrac{\beta_{\textsc{s},\textsc{l}}\omega_{\textsc{l}}}{2}\Bigr)}{\dfrac{\omega_{\textsc{h}}}{2}\Bigl(\coth\dfrac{\beta_{\textsc{s},\textsc{h}}\omega_{\textsc{h}}}{2}-\coth\dfrac{\beta_{\textsc{s},\textsc{l}}\omega_{\textsc{l}}}{2}\Bigr)}
\end{equation}
with the help of \eqref{E:fmsgsmfg}. It shows the dependence of the squeeze parameter and the bath temperatures in a nontrivial way.  For example, in the extreme squeezing limit $\eta\to\infty$, the efficiency $\xi^{(\textsc{sc})}$ reduces to
\begin{equation}
	\lim_{\eta\to\infty}\xi^{(\textsc{sc})}=\Bigl(1-\frac{\omega_{\textsc{l}}}{\omega_{\textsc{h}}}\Bigr)\frac{\omega_{\textsc{h}}+\omega_{\textsc{l}}}{2\omega_{\textsc{h}}}<\xi^{(\textsc{ad})}\,.
\end{equation}
Between these two  limits,  numerical means are needed to explore the full parameter space of the efficiency. {That is, in general $E_{\textsc{b}}$ will take the form
\begin{equation}
	E_{\textsc{b}}=\cosh2\eta_{\textsc{s}}\times\frac{\omega_{\textsc{h}}}{2}\,\coth\frac{\beta_{\textsc{l}}\omega_{\textsc{l}}}{2}
\end{equation}
with again $\eta_{\textsc{s}}$ computed via \eqref{E:fgbkjsbgdf} by the numerical approach.

\section{Conclusion}
We have investigated  some foundational theoretical issues for the isentropic and the isothermal phases of continuous variables quantum Otto engines, in the nonMarkovian regimes based on the conceptual and technical framework of the Hu-Paz-Zhang master equation.  These aspects are shown: 1) a reminder that  results based on the Lindblad or Caldeira-Leggett equations at {\it very low temperatures} are unreliable,  2) pointing out some possible flaws in the literature concerning the {\it late time squeezed thermal state};  3) exploring {\it strong system-bath coupling} in the isothermal phases, as well as  4) exploring  {\it nonadiabatic frequency variation} in the isentropic phases,  and 5) presenting the proper  {\it junction conditions} between the isentropic and the isothermal phases.  As noted earlier,  by crutinizing the validity of assumptions and approximations in existing theories and in exploring a broader range of parameter spaces in the open quantum systems and squeezed quantum systems, we hope to provide a more solid theoretical foundation for the search for optimal quantum engine performance, and more general quantum thermodynamic applications of these systems.\\

\noindent{\bf Acknowledgment}  OA acknowledges a GRA from MCFP in 2021 and summer support from the Joint Quantum Institute in 2020. JTH is supported by the Ministry of Science and Technology of Taiwan under Grant No.~MOST 110-2811-M-008-522.

\newpage

\appendix

\section{Late-time coefficients of HPZ master equation\label{coeff}}
The HPZ master equation for a bath of squeezed harmonic oscillators is derived in~\cite{HM94} for the convenience of readers following that stream of work \cite{HKM94,KMH97}:
\begin{align}\label{me}
	i\,\frac{\partial\hat{\rho}}{\partial t} &= \bigl[\hat{H}_{\text{ren}},\hat{\rho}\bigr] + i\,D_{pp}\bigl[\hat{\chi},\bigl[\hat{\chi},\hat{\rho}\bigr]\bigr] + i\,D_{xx}\bigl[\hat{p},\bigl[\hat{p},\hat{\rho}\bigr]\bigr]+	 i\,(D_{px}+D_{xp})\bigl[\hat{\chi},\bigl[\hat{p},\hat{\rho}\bigr]\bigr] + \Gamma\bigl[\hat{\chi},\bigl\{\hat{p},\hat{\rho}\bigr\}\bigr]\,,\notag \\
\hat{H}_{\text{ren}} &= \frac{\hat{p}^2}{2m}+\frac{m\omega^2\hat{\chi}^2}{2}\,,\qquad\qquad\qquad\omega^2 = \omega_{\text{b}}^2 -\int\!d\kappa\; \frac{I(\kappa)}{\kappa}\,,
\end{align}
where $\omega_{\text{b}}$ is the bare frequency of the system and $I(\omega)$ is the spectral density of the bath.

The master equation is derived using the influence functional formalism and the evolution operator can be expressed with a function of initial and final values of $\Delta$, $\Sigma$ functions of forward and backward classical trajectories multiplied by a path-independent factor. The classical trajectories of $\Delta$ and $\Sigma$ functions can be reduced to a linear combination of initial and final values with coefficients $u_1$, $u_2$, $v_1$, $v_2$  satisfying appropriate boundary conditions. For an Ohmic bath with spectral density
\begin{equation}
	I(\kappa)=\frac{2m\gamma\kappa \Theta(\Lambda - \kappa)}{\pi}\,,
\end{equation}
where $\Lambda$ is a cutoff frequency much larger than all relevant frequency scales of the system so that the dissipation is time local but the renormalized frequency remains positive, these functions are given by
\begin{align}
	\omega_{\text{r}}^2 &= \omega^2 - \gamma^2\,, \\
	u_1(s,t)&=-\frac{\sin[\omega_{\text{r}}(s-t)]\,e^{-\gamma s}}{\sin(\omega_{\text{r}}t)}\,,&u_2(s,t)&=\frac{\sin(\omega_{\text{r}}s)\,e^{-\gamma (s-t)}}{\sin(\omega_{\text{r}}t)}\,,\\
	v_{1}(s,t)&=u_{2}(t-s,t)\,,&v_{2}(s,t)&=u_{1}(t-s,t)\,.
\end{align}
Ref. \cite{HM94} defines functions $b_1(t)$, $\dots$, $b_4(t)$ and $a_{ij}(t)$, {where $i$ and $j$ take the values of $1$ and $2$}, in terms of $u$ and $v$ to be used in the expressions of the master equation coefficients
\begin{align}
	b_1(t)&=-b_4(t)=m\Bigl[\omega_{\text{r}}\cot(\omega_{\text{r}}t)-\gamma\Bigr]\,,\\
	b_2(t)&=\frac{m\omega_{\text{r}}\,e^{\gamma t}}{\sin(\omega_{\text{r}}t)}\,,\\
	b_3(t)&=-\frac{m\omega_{\text{r}}\,e^{-\gamma t}}{\sin(\omega_{\text{r}}t)}\,,\\
	a_{ij}(t)&=\frac{1}{1+\delta_{ij}} \int_{0}^{t}\!ds\int_{0}^{t}\!ds'\; v_i(s)\nu(s,s')v_j(s')\,,\\
\nu(s,s')&= \int_0^{\infty}\!d\kappa\;I(\kappa)\,\coth\frac{\beta\kappa}{2}\,\Bigl\{\cosh2\eta\cos[\kappa(s-s')]-\sinh2\eta\cos[2\theta-\kappa(s+s')]\Bigr\} \,.
\end{align}
Notice that we set all of the bath oscillators with the same squeezing parameters $\eta$ and $\theta$ and temperature $\beta^{-1}$ for simplicity.

At this point, we need to give an expression for $a_{ij}$ to evaluate the coefficients of the master equation. However, the integral over spectral density in the expression of $\nu(s,s')$ cannot be solved analytically, so we will introduce its Laplace transform $\tilde{\nu}(\sigma,\sigma')$ to be able to deal with the time integrals in the expression of $a_{ij}$ first and then take the integral over the spectral density numerically. It is straightforward to check that any straight line in the complex plane with fixed real part such that $\operatorname{Re}(\sigma)$, $\operatorname{Re}(\sigma')>0$ is in the region of convergence once we impose ${v_{1}(t<0)=v_{2}(t<0)=0}$ and we stick to this choice for our calculation. The Laplace transform of $\nu(s,s')$ is given by
\begin{align}
	\tilde{\nu}(\sigma,\sigma') &= \int_0^{\infty}\!d\kappa\; \frac{S(\kappa)}{(\sigma^2+\kappa^2)(\sigma'^{2}+\kappa^2)} \Bigl\{\cosh2\eta\times(\sigma\sigma'+\kappa'^{2})\Bigr.\\
	&\qquad\qquad\qquad\qquad\qquad -\Bigl.\sinh2\eta\bigl[(\sigma\sigma'-\kappa^{2})\cos2\theta+\kappa(\sigma+\sigma')\sin2\theta\bigr]\Bigr\}\,,\notag
\end{align}
where $S(\kappa) = \dfrac{2\gamma m \kappa \Theta(\Lambda - \kappa)}{\pi}\coth\dfrac{\beta\kappa}{2}$. Using this expression, we can bring the expression of $a_{ij}$ to the form
\begin{align}
	a_{ij}(t)&=-\frac{1}{4\pi^2(1+\delta_{ij})} \int_{\epsilon-i\infty}^{\epsilon+i\infty}\!d\sigma\int_{\epsilon-i\infty}^{\epsilon+i\infty}\!d\sigma'\int_{0}^{t}\!ds\int_{0}^{t}\!ds'\; v_i(s)\tilde{\nu}(\sigma,\sigma')v_j(s')\,e^{\sigma s+\sigma's'}\,,
\label{a-lt}
\end{align}
where $\epsilon$ is an arbitrary real number strictly greater than zero. Plugging in the expressions for $v_{1}(t)$, $v_{2}(t)$ in Eq.~\eqref{a-lt} and taking the integrals over time, we obtain the following expressions for $a_{ij}$,
\begin{align}
	a_{11}(t)&=-\frac{1}{8\pi^2\sin^2(\omega_{\text{r}}t)} \int_{\epsilon-i\infty}^{\epsilon+i\infty}\!d\sigma\int_{\epsilon-i\infty}^{\epsilon+i\infty}\!d\sigma'\;\tilde{\nu}(\sigma,\sigma') \nonumber \\
&\qquad\qquad\qquad\qquad\qquad\times\frac{\omega_{\text{r}}e^{(\gamma+\sigma)t}-(\gamma+\sigma)\sin(\omega_{\text{r}}t)-\omega_{\text{r}}\cos(\omega_{\text{r}}t)}{(\gamma+\sigma)^2 + \omega_{\text{r}}^2} \nonumber \\
&\qquad\qquad\qquad\qquad\qquad\times \frac{\omega_{\text{r}}e^{(\gamma+\sigma')t}-(\gamma+\sigma')\sin(\omega_{\text{r}}t)-\omega_{\text{r}}\cos(\omega_{\text{r}}t)}{(\gamma+\sigma')^2 + \omega_{\text{r}}^2}\,, \\
	a_{12}(t)&= -\frac{\dot{a}_{11}}{\dot{v}_1(t)} = \frac{\dot{a}_{11}\sin(\omega_{\text{r}}t)}{\omega_{\text{r}}\,e^{\gamma t}}\,,\qquad\qquad\qquad\qquad \text{  from Eq.~(D5) of Ref.~\cite{HM94}}\,,\\
	a_{22}(t)&=-\frac{1}{8\pi^2\sin^2(\omega_{\text{r}}t)}\int_{\epsilon-i\infty}^{\epsilon+i\infty}\!d\sigma\int_{\epsilon-i\infty}^{\epsilon+i\infty}d\sigma'\;\tilde{\nu}(\sigma,\sigma') \nonumber \\
&\qquad\qquad\qquad\qquad\qquad\times\frac{\omega_{\text{r}}\,e^{-\gamma t}+[(\gamma+\sigma)\sin(\omega_{\text{r}}t)-\omega_{\text{r}}\cos(\omega_{\text{r}}t)]\,e^{\sigma t}}{(\gamma+\sigma)^2 + \omega_{\text{r}}^2} \nonumber \\
&\qquad\qquad\qquad\qquad\qquad\times \frac{\omega_{\text{r}}\,e^{-\gamma t}+[(\gamma+\sigma')\sin(\omega_{\text{r}}t)-\omega_{\text{r}}\cos(\omega_{\text{r}}t)]\,e^{\sigma' t}}{(\gamma+\sigma')^2 + \omega_{\text{r}}^2}\,.
\end{align}
These integrals are evaluated using the residue theorem and an appropriate closed integration contour including the line from $\epsilon-i\infty$ to $\epsilon-i\infty$ and the results are given in Eqs.~\eqref{a11}, \eqref{a22} and \eqref{a12} with the exponentially decaying terms omitted thanks to late-time assumption,
\begin{align}
	a_{11}(t\gg1/\gamma)&= \int_0^{\infty}\!d\kappa\;\frac{S(\kappa)\,\omega_{\text{r}}^2\,e^{2\gamma t} }{2\sin^2(\omega_{\text{r}}t)} \biggl[\frac{\cosh2\eta}{|(\gamma+i\kappa)^2 + \omega_{\text{r}}^2|^2}\biggr. \notag \\
			&\qquad\qquad\qquad\qquad\qquad\qquad-\biggl.\sinh2\eta\,\operatorname{Re}\biggl\{\frac{e^{2i(\kappa t - \theta)}}{[(\gamma+i\kappa)^2 + \omega_{\text{r}}^2]^2}\biggr\}\biggr] \,, \label{a11}\\
	a_{22}(t\gg1/\gamma)&= \int_0^{\infty}\!d\kappa\;\frac{S(\kappa)}{2}\,\biggl[\frac{\cosh2\eta\,[(\gamma-\omega_{\text{r}}\cot\omega_{\text{r}}t)^2+\kappa^2]}{|(\gamma+i\kappa)^2 + \omega_{\text{r}}^2|^2}\biggr. \nonumber \\
&\qquad\qquad\qquad- \sinh2\eta\,[(\gamma-\omega_{\text{r}}\cot\omega_{\text{r}}t)^2-\kappa^2]\,\operatorname{Re}\biggl\{\frac{e^{2i(\kappa t - \theta)}}{[(\gamma+i\kappa)^2 + \omega_{\text{r}}^2]^2}\biggr\}\nonumber \\
&\qquad\qquad\qquad-\biggl. 2\kappa(\gamma-\omega_{\text{r}}\cot\omega_{\text{r}}t)\,\operatorname{Im}\biggl\{\frac{e^{2i(\kappa t - \theta)}}{[(\gamma+i\kappa)^2 + \omega_{\text{r}}^2]^2} \biggr\}\biggr] \,,\label{a22}\\
	a_{12}(t\gg1/\gamma)&= \int_0^{\infty}\!d\kappa\;\frac{S(\kappa)\,\omega_{\text{r}}\,e^{\gamma t}}{\sin(\omega_{\text{r}} t)}  \biggl[(\gamma-\omega_{\text{r}}\cot\omega_{\text{r}}t)\biggl(\frac{\cosh2\eta}{|(\gamma+i\kappa)^2 + \omega_{\text{r}}^2|^2}\biggr.\biggr. \nonumber \\
&\qquad\qquad\qquad\qquad\qquad\qquad-\biggl. \sinh2\eta\,\operatorname{Re}\biggl\{\frac{e^{2i(\kappa t - \theta)}}{[(\gamma+i\kappa)^2 + \omega_{\text{r}}^2]^2}\biggr\} \bigg)\nonumber \\
&\qquad\qquad\qquad\qquad\qquad\qquad+\biggl. \kappa\sinh2\eta\,\operatorname{Im}\biggl\{\frac{e^{2i(\kappa t - \theta)}}{[(\gamma+i\kappa)^2 + \omega_{\text{r}}^2]^2} \biggr\}\biggr]\,.\label{a12}
\end{align}
Having explicit expressions for $b_1$, $\cdots$, $b_4$ and $a_{ij}$, and using the equations~(D8) to~(D11) from Ref.~\cite{HM94}, the late-time coefficients of Eq.~\eqref{me} are the following. The integration $\displaystyle\int_0^{\infty}\!d\kappa\;S(\kappa)\cdots $ is omitted but implied in the expressions of $D_{pp}$ and $D_{px}$ for a cleaner notation,
\begin{align}
	\Gamma &= \gamma\,, \qquad\qquad\qquad\qquad\qquad D_{xx} = 0\,,\label{gamma}\\
	-D_{xp} &= -D_{px} = \frac{(\kappa^2 - \omega^2)\,\cosh2\eta}{2m|(\gamma+i\kappa)^2 + \omega_{\text{r}}^2|^2}+ \frac{\sinh2\eta}{2m}\biggl[(3\kappa^2+\omega^2)\operatorname{Re}\biggl\{\frac{e^{2i(\kappa t - \theta)}}{[(\gamma+i\kappa)^2 + \omega_{\text{r}}^2]^2}\biggr\} \biggr.\nonumber \\
	&\qquad\qquad\qquad\qquad\qquad\qquad\qquad\qquad\qquad-\biggl. 2\kappa\gamma \operatorname{Im}\biggl\{\frac{e^{2i(\kappa t - \theta)}}{[(\gamma+i\kappa)^2 + \omega_{\text{r}}^2]^2}\biggr\}\biggr]\,, \\
	D_{pp} &= \frac{-2\gamma\kappa^2\,\cosh2\eta}{|(\gamma+i\kappa)^2 + \omega_{\text{r}}^2|^2} + \sinh2\eta\, \biggl[(\kappa^2-\omega^2)\kappa\,\operatorname{Im}\biggl\{\frac{e^{2i(\kappa t - \theta)}}{[(\gamma+i\kappa)^2 + \omega_{\text{r}}^2]^2}\biggr\}\biggr.\notag\\
	 &\qquad\qquad\qquad\qquad\qquad\qquad\qquad\qquad\qquad-\biggl.  6\gamma\kappa \operatorname{Re}\biggl\{\frac{e^{2i(\kappa t - \theta)}}{[(\gamma+i\kappa)^2 + \omega_{\text{r}}^2]^2}\biggr\}\biggr]\,.
\end{align}
Before proceeding to find the steady state of Eq.~\eqref{me}, we can make a further approximation for $D_{xp}$ and $D_{pp}$. For a large enough time $t$, we can assume the rotating terms of the integrand behave smoothly enough with changes of $\kappa$ in the order of $1/t$ so that the integral over bath spectrum becomes negligible. Eliminating these terms give the following expressions\footnote{Notice that the signs of $D_{xp}$, $D_{px}$ terms differ in Eqs. (3.4) and (D6) of Ref.~\cite{HM94}. Eqs.~\eqref{gamma} and \eqref{dpp} refers to the coefficients of Eq.~\eqref{me}.} for $D_{xp}$ and $D_{pp}$
\begin{align}
	D_{xp} &=\frac{(\omega^2 - \kappa^2)\,\cosh2\eta}{2m\,|(\gamma+i\kappa)^2 + \omega_{\text{r}}^2|^2}\,, &D_{pp} &= -\frac{2\gamma\kappa^2 \cosh2\eta}{|(\gamma+i\kappa)^2 + \omega_{\text{r}}^2|^2}\,. \label{dpp}
\end{align}

\newpage

\begin{thebibliography}{999}

\bibitem{QTDbooks} 
	J. Gemmer, M. Michel, G. Mahler, {\sl Quantum Thermodynamics} (Springer Verlag,  New York, 2009). 

	R. Kosloff,  {\it Quantum thermodynamics: A dynamical viewpoint}, Entropy {\bf15}, 2100 (2013).

	S. Vinjanampathy, J. Anders,  {\it Quantum thermodynamics}, Contemp. Phys. {\bf57}, 545 (2016).

\bibitem{KosLev} 
	R. Kosloff and A. Levy, {\it Quantum heat engines and refrigerators: Continuous devices}, Annu. Rev. Phys. Chem. {\bf65}, 365 (2014). 

\bibitem{Alicki} 
	A. Levy, R. Alicki and R. Kosloff, {\it Quantum refrigerators and the third law of thermodynamics}, Phys. Rev. E {\bf85}, 061126 (2012) .

\bibitem{small} 
	N. Linden, S. Popescu and P. Skryzypczyk, {\it How small can thermal machines be? The smallest possible refrigerator}, Phys. Rev. Lett. {\bf 105}, 130401 (2010).
		
\bibitem{MarPaz} 
	E. A. Martinez and J. P. Paz, {\it Dynamics and thermodynamics of linear quantum open systems},  Phys. Rev. Lett. {\bf110}, 130406 (2013).

\bibitem{Uzdin}
	R. Uzdin, A. Levy and R. Kosloff, {\it Quantum heat machine equivalence, work extraction beyond Markovianity, and strong coupling via heat exchangers},  Entropy {\bf18}, 124 (2016).

\bibitem{Bre16}  
	H.-P. Breuer, E.-M. Laine, J. Piilo, and B. Vacchini, {\it Non-Markovian dynamics in open quantum systems}, Rev. Mod. Phys. {\bf88}, 021002 (2016).

\bibitem{DeV17}
	I. de Vega and D. Alonso, {\it Dynamics of non-Markovian open quantum systems}, Rev. Mod. Phys. {\bf89}, 015001 (2019).
		
\bibitem{HuePle} 
	D. Tamascelli, A. Smirne, S. F. Huelga and M. B. Plenio, {\it Nonperturbative treatment of non-Markovian dynamics of
open quantum systems}, Phys. Rev. Lett. {\bf120}, 030402 (2018).

\bibitem{QTD1} 
	J.-T. Hsiang, C. H. Chou, Y. Suba{\c s}{\i} and B. L. Hu, {\it Quantum thermodynamics from the nonequilibrium dynamics of open systems: Energy, heat capacity, and the third law}, Phys. Rev. E {\bf97}, 012135 (2018).
	
\bibitem{Klaers}
		J. Klaers, S. Faelt, A. \.{I}mamo{\u g}lu and E. Togan, {\it Squeezed thermal reservoirs as a resource for a nanomechanical engine beyond the Carnot limit}, Phys. Rev. X {\bf7}, 031044 (2017).

\bibitem{UW12}
		U. Weiss, 
		{\textsl{Quantum Dissipative Systems, 4th Edition}} (World Scientific, Singapore, 2012).

\bibitem{BrePet}  
 		H. P. Breuer, and F. Petruccione, 
		{\textsl{The Theory of Open Quantum Systems, 2nd Edition}} (Oxford University Press, Oxford, 2007).	
		
\bibitem{RivHue}
		A. Rivas, and  S. F. Huelga, 
		{\textsl{Open quantum systems: An Introduction}} (Springer, Berlin, Heidelberg, 2012).
		
\bibitem{CalHu88} 
		E. Calzetta, and B. L. Hu, 
		{\textit{Nonequilibrium quantum fields: Closed-time-path effective action, Wigner function and Boltzmann equation}}, Phys. Rev. D \textbf{37}, 2878 (1988).
		
\bibitem{CalHu08} 
		E. Calzetta, and B. L. Hu, 
		{\textsl{Nonequilibrium Quantum Field Theory}} (Cambridge University Press, Cambridge, 2008). 
	
\bibitem{JR09}
		J. Rammer, 
		{\textsl{Quantum Field Theory of Non-equilibrium States}} (Cambridge University Press, Cambridge, 2009). 
		
\bibitem{AK11}
		A. Kamenev, 
		{\textsl{Field Theory of Non-Equilibrium Systems}} (Cambridge University Press, Cambridge, 2011).
		
\bibitem{Berges} 
		J. Berges, 
		{\textit{Nonequilibrium Quantum fields: From cold atoms to cosmology}}, in Lecture Notes of the Les Houches Summer School, Vol. 99, \textsl{Strongly Interacting Quantum Systems out of Equilibrium} (Oxford University Press, Oxford, 2016); 
		{[arxiv:1503.02907]}. 

\bibitem{Ross} 
	J. Ro{\ss}nagel, S. T. Dawkins, K. N. Tolazzi, O. Abah, E. Lutz, F. Schmidt-Kaler and K. Singer, {\it A single-atom heat engine}, Science {\bf352}, 325 (2016).

\bibitem{Abah} 
	O. Abah, J. Ro{\ss}nagel, G. Jacob, S. Deffner, F. Schmidt-Kaler, K. Singer and E. Lutz, {\it Single ion heat engine with maximum efficiency at maximum power}, Phys. Rev. Lett. {\bf109} 203006 (2012).

\bibitem{KosRez} 
	R. Kosloff  and Y. Rezek,  {\it The quantum harmonic Otto cycle}, Entropy {\bf19}, 136 (2017).

\bibitem{DeffLutz} 
	S. Deffner, O. Abah, and E. Lutz, {\it Quantum work statistics of linear and nonlinear parametric oscillators}, Chem. Phys. {\bf375}, 200 (2010).

\bibitem{AbahLutz} 
	O. Abah and E. Lutz, {\it Efficiency of heat engines coupled to nonequilibrium reservoirs}, Europhys. Lett. {\bf106}, 20001 (2014).

\bibitem{Manzano} 
	G. Manzano,  F. Galve,  R. Zambrini  and J. M. R. Parrondo, {\it Entropy production and thermodynamic power of the squeezed thermal reservoir}, Phys. Rev.  E {\bf93}, 052120 (2016). 

\bibitem{Reid} 
	B. Reid, S. Pigeon, M. Antezza and G. De Chiara, {\it A self-contained quantum harmonic engine}, Europhys. Lett. {\bf120}, 60006 (2017).   

\bibitem{Niedensu}  
	W. Niedenzu, V. Mukherjee, A. Ghosh, A. G. Kofman and G. Kurizki, {\it Quantum engine efficiency bound beyond the second law of thermodynamics}, Nat. Commun. {\bf9}, 165 (2018).  

\bibitem{Pezzutto} 
	M. Pezzutto, M. Paternostro and Y. Omar, {\it An out-of-equilibrium non-Markovian quantum heat engine}, Quantum Sci. Technol. {\bf4} 025002 (2019). 

\bibitem{ColMod}
	F. Ciccarello, G. M. Palma, and V. Giovannetti, {\it Collision-model-based approach to non-Markovian quantum dynamics}, Phys. Rev. A {\bf87}, 040103 (2013). 

	R. McCloskey and M. Paternostro, {\it Non-Markovianity and System-Environment Correlations in a microscopic collision model},  Phys. Rev. A {\bf89}, 052120 (2014).

	S. Lorenzo, F. Ciccarello, and G. M. Palma, {\it Class of exact memory-kernel master equations},  Phys. Rev. A {\bf93}, 052111 (2016). 

	S. Kretschmer, K. Luoma and W. T. Strunz,  {\it Collision model for non-Markovian quantum dynamics}, Phys. Rev. A {\bf94}, 012106  (2016).

	M. Pezzutto, M. Paternostro and Y. Omar, {\it Implications of non-Markovian dynamics for the Landauer bound}, New J. Phys. {\bf18}, 123018 (2016).

	B. \c{C}akmak, M. Pezzutto, M. Paternostro and {\"O}. E. M\"ustecapl{\i}o{\u g}lu, {\it Non-Markovianity, coherence, and system-environment correlations in a long-range collision model}, Phys. Rev. A {\bf96}, 022109 (2017).

	S. Lorenzo, F. Ciccarello, G. M. Palma and B. Vacchini, {\it Quantum non-Markovian piecewise dynamics from collision models}, Open Syst. Inf. Dyn. {\bf24} 1740011 (2017).

	S. Campbell, F. Ciccarello, G. M. Palma and B. Vacchini,  {\it System-environment correlations and Markovian embedding of quantum non-Markovian dynamics}, Phys. Rev. A {\bf98} 012142 (2018).

	A. Pozas-Kerstjens, E. G. Brown and K. V. Hovhannisyan, {\it A quantum Otto engine with finite heat baths: energy, correlations, and degradation},  New J. Phys. {\bf20} 043034 (2018).
 
	M. Pezzutto, M. Paternostro and Y. Omar, {\it An out-of-equilibrium non-Markovian quantum heat engine}, Quantum Sci. Technol. {\bf4} 025002 (2019).

	G. De Chiara and M. Antezza,  {\it Quantum machines powered by correlated baths}, Phys. Rev. Res. {\bf2}, 033315 (2020). 

	V. Singh and {\"O}. E. M\"ustecapl{\i}o{\u g}lu, {\it Performance bounds of non-adiabatic quantum harmonic Otto engine and refrigerator under a squeezed thermal reservoir}, Phys. Rev. E {\bf102}, 062123  (2020). 

\bibitem{AHH2}  
	O. Ar{\i}soy, J.-T. Hsiang and B. L. Hu,  {\it Quantum Heat Engine in nonMarkovian squeezed baths:  Collision theory versus Brownian models}, in preparation.
 
\bibitem{Lindblad} 
	G. Lindblad, {\it On the generators of quantum dynamical semigroups}, Commun. Math. Phys. {\bf48}, 119 (1976).

\bibitem{GKS} 
	V. Gorini, A. Kossakowski, E. C. G. Sudarshan,  {\it Completely positive dynamical semigroups of $N$-level systems}, J. Math. Phys. {\bf17}, 821 (1976).

\bibitem{BornMark} 
	D. F. Walls and G. J. Milburn, {\sl Quantum Optics} (Springer Verlag, 2007).

\bibitem{CalLeg}  
	A. O. Caldeira and A. J. Leggett, {\it Path integral approach to quantum Brownian motion}, Phys. A (Amsterdam) {\bf121}, 587 (1983).

\bibitem{AmbBun} 
	V. Ambegaokar, {\it Quantum Brownian motion and its classical limit}, Phys. Chem. {\bf95}, 400 (1991).

\bibitem{Haake} 
	F. Haake and R. Reibold, {\it Strong damping and low-temperature anomalies for the harmonic oscillator}, Phys. Rev. A {\bf32}, 2462 (1985). 

\bibitem{Garraway} 
	B. M. Garraway, {\it Nonperturbative decay of an atomic system in a cavity}, Phys. Rev. A {\bf55}, 2290 (1997).

\bibitem{Maciej}
	A. Lampo, S. H. Lim, J. Wehr, P. Massignan, and M. Lewenstein, {\it Lindblad model of quantum Brownian motion}, Phys. Rev. A {\bf94}, 042123 (2016).

\bibitem{Jacobs} 
	G. McCauley, B. Cruikshank, D. I. Bondar and K. Jacobs, {\it Accurate Lindblad-form master equation for weakly damped quantum systems across all regimes}, NPJ Quantum Information {\bf6}, 74 (2020).

\bibitem{HPZ92} 
	B. L. Hu, J. P. Paz and Y. Zhang, {\it Quantum Brownian motion in a general environment: exact master equation with nonlocal dissipation and colored noise}, Phys. Rev. D {\bf45}, 2843 (1992).

\bibitem{HPZ93} 
	B. L. Hu, J. P. Paz and Y. Zhang, {\it Quantum Brownian motion in a general environment. II. Nonlinear coupling and perturbative approach}, Phys. Rev. D {\bf47}, 1576 (1993).

\bibitem{HalYu96} 
	J. J. Halliwell and T. Yu. {\it Alternative derivation of the Hu-Paz-Zhang master equation of quantum Brownian motion}, Phys. Rev. D {\bf53}, 2012 (1996).

\bibitem{HM94} 
	B. L. Hu and A. Matacz, {\it Quantum Brownian motion in a bath of parametric oscillators: A model for system-field interactions}, Phys. Rev.  D {\bf49}, 6612 (1994).
     
\bibitem{HKM94}	
	B. L. Hu, G. Kang and A. Matacz,  {\it Squeezed vacua and the quantum statistics of cosmological particle creation}, Int. J. of Mod. Phys. A {\bf9}, 991 (1994).

\bibitem{RHK97}  
	A. Raval, B. L. Hu and Don Koks, {\it Near-thermal radiation in detectors, mirrors, and black holes: A stochastic approach}, Phys. Rev.  D {\bf55}, 4795 (1997).

\bibitem{KMH97} 
	D. Koks, A. Matacz and B. L. Hu, {\it Entropy and uncertainty of squeezed quantum open systems}, Phys. Rev.  D {\bf55}, 5917 (1997).

\bibitem{HHCosSq}   
	J.-T. Hsiang and B. L. Hu, {\it Dynamical response of an Unruh-DeWitt detector in a quantum field over the history of the Universe}, in {\sl Universe}: Special issue on  "Quantum Aspects of the Universe", in preparation.

\bibitem{FDRSq}   
	J.-T. Hsiang and B. L. Hu,  {\it Fluctuation-dissipation relation for a quantum Brownian parametric oscillator in a squeezed thermal bath}, in preparation for Phys. Rev. D.  

\bibitem{NEqFE}  
	J.-T. Hsiang and B. L. Hu,  {\it Nonequilibrium quantum free energy and effective temperature, generating functional and influence action}, Phys. Rev. D {\bf103}, 065001 (2021);  [arXiv:2011.10468]. 

\bibitem{HHHMF}  
	J.-T. Hsiang and B. L. Hu,  {\it Nonequilibrium quantum thermodynamics at strong coupling: Hamiltonian of mean force and internal energy}, in preparation.

\bibitem{MarEmb} 
	P. Siegle, I. Goychuk, P. Talkner and P. H\"anggi, {\it Markovian embedding of non-Markovian superdiffusion}, Phys. Rev. E {\bf81}, 011136 (2010).

\bibitem{QCont} 
	H. M. Wiseman and G. J. Milburn,  {\sl Quantum Measurement and Control} (Cambridge University Press, Cambridge,  2009).

\bibitem{HK53}
	K. Husimi, {\it Miscellanea in elementary quantum mechanics, II}, Prog. Theor. Phys. {\bf9}, 381 (1953).

\bibitem{STA}
	D. Gu\'ery-Odelin, A. Ruschhaupt, A. Kiely, E. Torrontegui, S. Mart{\'i}nez-Garaot and J. G. Muga {\it Shortcuts to adiabaticity : concepts, methods, and applications}, Rev. Mod. Phys. {\bf 91}, 045001 (2019).
\end{thebibliography}

\end{document}